\documentclass[paper]{JHEP3} 

\JHEPspecialurl{http://jhep.sissa.it/JOURNAL/JHEP3.tar.gz}

\usepackage{epsfig,multicol}
\usepackage{feynmf}

\newcommand\fverb{\setbox\pippobox=\hbox\bgroup\verb}
\newcommand\fverbdo{\egroup\medskip\noindent%
			\fbox{\unhbox\pippobox}\ }
\newcommand\fverbit{\egroup\item[\fbox{\unhbox\pippobox}]}
\newcommand {\beq}{\begin{equation}}
\newcommand {\eeq}{\end{equation}}
\newcommand {\beqa}{\begin{eqnarray}}
\newcommand {\eeqa}{\end{eqnarray}}
\newcommand {\n}{\nonumber \\}

\newcommand {\ee}{\mbox{e}}

\newcommand {\dd}{\mbox{d}}

\newcommand {\del}{\partial}

\newcommand {\defeq}{\stackrel{\rm def}{=}}

\newcommand {\latsp}{a}

\def\fsl#1{{#1}\!\!\! / }

\newbox\pippobox

\title{Lattice Perturbation Theory in\\
Noncommutative Geometry and\\
Parity Anomaly in 3D Noncommutative QED}

\author{Jun Nishimura\\
Department of Physics, Nagoya University\\
Furo-cho, Chikusa-ku, Nagoya 464-8602, Japan\\
E-mail: \email{nisimura@eken.phys.nagoya-u.ac.jp}}
\author{Miguel A. V\'azquez-Mozo\\
Theory Division, CERN\\
CH-1211 Geneva 23, Switzerland\\
E-mail: \email{Miguel.Vazquez-Mozo@cern.ch}}

\preprint{\heplat{0210017}\\
DPNU-02-32\\
CERN-TH/2002-254}	

\abstract{We formulate lattice perturbation theory for
gauge theories in noncommutative geometry.
We apply it to three-dimensional noncommutative QED
and calculate the effective action induced by Dirac fermions.
In particular ``parity invariance'' of a massless theory
receives an anomaly expressed by the noncommutative Chern-Simons
action. The coefficient of the anomaly is labelled
by an integer depending on the lattice action, which is a 
noncommutative counterpart of the phenomenon 
known in the commutative theory.
The parity anomaly can also be obtained using Ginsparg-Wilson fermions,
where 
the masslessness is guaranteed at finite lattice spacing.
This suggests a natural definition of 
the lattice-regularized Chern-Simons theory 
on a noncommutative torus, which could enable nonperturbative studies
of quantum Hall systems.
}


\begin{document} 


\section{Introduction}

Quantum field theory on noncommutative spaces has been a subject of much 
activity in recent years 
(see \cite{dougnek,Szabo:2001kg} for comprehensive reviews).
This attention was originally triggered 
by its intimate relationship with string theory, but the study 
of noncommutative field theories has interest in its own. 
From a mathematical physics point of view 
they provide us with a class of nonlocal quantum field theories which 
nonetheless seem to be well defined.
Also, since noncommutative field theories are essentially
theories of dipoles \cite{dipoles} 
they can be also useful in the analysis of systems with
dipolar excitations in condensed matter physics.

In this context, noncommutative Chern-Simons (NCCS) theory 
in $(2+1)$-dimensions 
is specially appealing 
because of its applications to quantum Hall systems \cite{susskind,hallCS}.
Here the noncommutativity is introduced only in the spatial direction
and the resulting deformation of the ordinary gauge invariance into 
``star-gauge invariance'' is essential for the description of
the system which has the area preserving diffeomorphism 
invariance \cite{susskind}.
Remarkably, a finite-$N$ matrix model 
of NCCS theory is found to have physical states
which have one-to-one correspondence 
with Laughlin-type wave functions \cite{Hellerman:2001rj}.
It has also been pointed out that instabilities 
of the NCCS theory can describe 
the transition to the Wigner crystal, where spontaneous breakdown
of translational invariance is caused 
by the noncommutativity \cite{Barbon:2001dw}.

In studying noncommutative field theories
it is often useful to consider its matrix-model description.
This is reminiscent of its string/M-theoretic connections \cite{CDS,Aoki:1999vr}.
Here the space-time degrees of freedom and the internal (``color")
degrees of freedom are treated on equal footing, and
``star-gauge invariance" is simply described by the global U($\infty$)
symmetry which acts on the matrix indices.
The matrix model description is also useful for regularizing
noncommutative field theories \cite{Aoki:1999vr,
Bars:2000av,AMNS1,AMNS2,AMNS3},
since finite-$N$ twisted reduced models \cite{GO}
are interpreted as a lattice formulation of noncommutative 
field theories \cite{AMNS1,AMNS2,AMNS3}.
Such a lattice formulation provides the most reliable
method to study the quantum dynamics of noncommutative field theories 
in a fully nonperturbative manner.
Recently, the lattice regulatization has been applied to two-dimensional 
noncommutative Yang-Mills theory 
\cite{2dNCYM,Bietenholz:2002vj},
where (nonperturbative) renormalizability was demonstrated
for the first time in a noncommutative field theory (see also \cite{Profumo:2002cm}).
There the same theory was shown to have
an intriguing infrared property which may be described as
the Aharonov-Bohm effect with the magnetic field identified
with the inverse noncommutativity parameter.
The lattice formulation
has also been used to explore the phase diagram of 
noncommutative scalar field theories
\cite{Bietenholz:2002vj,Ambjorn:2002nj},
which is expected to be richer than in the commutative case,
as indicated by a self-consistent Hartree approximation 
\cite{Gubser:2000cd}.
In particular, as conjectured by Ref.\ \cite{Gubser:2000cd},
the ordered phase is found to split into
a uniformly ordered phase and a phase dominated by the stripe pattern.
In the latter phase, nonzero momentum modes acquire vacuum expectation
values, and therefore translational invariance
is spontaneously broken.

The aim of this paper is two-fold.
First we formulate perturbation theory for the 
lattice noncommutative gauge theory.
Although the most important virtue of 
the lattice regularization lies in its 
capability of nonperturbative studies,
it has also been used to clarify subtle issues
in perturbative aspects of gauge theories.
We consider this particularly important 
because the lattice construction of noncommutative chiral gauge
theories suggests a new mechanism of gauge anomaly
cancellation, which is not yet known 
in the continuum \cite{Nishimura:2001dq}.
As an application of the lattice perturbation theory,
we pick up a noncommutative version of three-dimensional QED, where 
the lattice calculation
indeed plays a crucial role in revealing peculiar properties
of the parity anomaly, given in terms of 
noncommutative Chern-Simons action.
The coefficient of the anomaly
is labelled by an integer depending on the lattice action,
which is a counterpart of the phenomenon observed by 
Coste and L\"uscher \cite{cl} in the commutative theory.
The commutative limit is smooth when the fermions 
are in the fundamental representation,
but it is {\em not} for fermions in the adjoint representation
due to a characteristic property of noncommutative field theories
known as the UV/IR mixing \cite{iruv}. 
As a special case our result includes Chu's result \cite{Chu}
obtained in the continuum by using the Pauli-Villars regularization.

Another aim of this work is to construct a lattice regularization
of NCCS theory, which has important
applications to quantum Hall systems as mentioned above.
In general, defining a Chern-Simons term on the lattice is
not straightforward due to its topological nature.
A recent proposal \cite{3dGW}
is to utilize Ginsparg-Wilson fermions 
in odd dimensions and to define the Chern-Simons term using the
parity violating part of the effective action induced by the fermion.
We show that this proposal works also in the noncommutative case.
As we mentioned above,
the parity anomaly in the three-dimensional noncommutative QED is given 
by the NCCS term in the 
continuum limit of the lattice theory we started with.
The same result can be obtained from Ginsparg-Wilson fermions,
where the masslessness is guaranteed at finite lattice spacing.
This suggests a natural definition of 
the lattice-regularized Chern-Simons theory 
on a noncommutative torus, which could enable nonperturbative studies
of quantum Hall systems.
In this regard, we recall that 
a finite-$N$ matrix model has been proposed
as a regularized description of NCCS theory 
on a cylinder \cite{polychr2}.
Although our theory can also be mapped to 
a finite-$N$ matrix model,
the two proposals appear to be quite different.

The rest of the paper is organized as follows.
In Section \ref{matrixmodel}, we review the lattice formulation
of noncommutative gauge theories and formulate
a perturbation theory based on Feynman rules.
In Section \ref{continuum}, we present calculations of the parity
anomaly in noncommutative QED.
In Section \ref{CSterm} we discuss the parity anomaly
using Ginsparg-Wilson fermions,
which leads to a proposal for
a lattice-regularized noncommutative Chern-Simon theory.
Section \ref{summary} is devoted to summary and discussions.
Finally, in Appendix \ref{a1} 
a calculation of the parity-violating terms in the 
effective action by the direct evaluation of the fermionic determinant
is presented, while Appendix \ref{A2} contains some details of the 
computation of the Feynman integrals involved in the diagrammatic
calculation of Section \ref{continuum}.

\setcounter{equation}{0}
\section{Lattice perturbation theory in noncommutative geometry 
}
\label{matrixmodel}

In this section we begin by reviewing the lattice formulation 
of noncommutative gauge theories
developed in Ref.\ \cite{AMNS1,AMNS2,AMNS3}
(see \cite{Makeenko:2000tc,Szabo:2001qd,yuri} for reviews).
In the literature it is common to start from a finite-$N$
matrix model, which is then shown to be equivalent to
the lattice formulation of a noncommutative field theory.
Indeed, the matrix model representation has proven useful for
numerical analyses
\cite{2dNCYM,Bietenholz:2002vj,Ambjorn:2002nj,Nishimura:2001dq}.
Here we will work directly with the lattice formulation
and derive the Feynman rules, which are used
in the perturbative evaluation of the effective action 
induced by fermions.
Although we will focus on
noncommutative QED in $d=2m+1$ dimensions,
the lattice perturbation theory can be formulated
for any other noncommutative field theories
in the same way.
%


\subsection{Noncommutative QED on the lattice}

In noncommutative geometry, space-time coordinates are 
treated as Hermitian operators obeying the commutation relation
$[ \hat{x}_\mu , \hat{x}_\nu ] = i \, \theta_{\mu\nu}$, where 
$\theta_{\mu\nu}$ is a real anti-symmetric matrix.
In $(2m+1)$ dimensions, due to a property of antisymmetric matrices,
coordinates can always be chosen
in such a way that one of them commutes with all the others,
resulting in 
\begin{eqnarray}
[\hat{x}_{i},\hat{x}_{j}]=i \, \theta_{ij} \ , 
\hspace*{1cm} [\hat{x}_{i}, \hat{x}_{d}]=0 \ , 
\hspace*{0.5cm} (i,j=1,\ldots,2m) \ ,
\end{eqnarray}
where, for simplicity, the $2m\times 2m$ noncommutative matrix $\theta_{ij}$
is taken to be of the form
\begin{eqnarray}
\theta_{ij}\equiv \theta \, \varepsilon_{ij} \ ,
\quad\quad
\varepsilon = 
\left(
\begin{array}{ccccccc}
0 & & -1 &  &  &  \\
1  &  &  0  &   &  &  \\
        &  &    & \ddots  &  &  &  \\
        &  &   &         & 0 & & -1 \\
        &  &   &         & 1 & & 0 
\end{array}
\right) \ .
\label{eps}
\end{eqnarray}
We regard the commuting coordinate $\hat{x}_{d}$ 
as the Euclidean time after the Wick rotation.
%
Field theories on a noncommutative geometry can be obtained by
replacing an ordinary field $\phi(x)$ by an operator $\phi(\hat{x})$.
An equivalent way to describe noncommutative field theories,
which we are going to use in what follows,
is to keep the ordinary field $\phi(x)$ but to replace
the ordinary product of fields, 
say $\phi_1(x)$ and $\phi_2(x)$, by the star-product
\beq
\phi_1(x) \star \phi_2(x) = \phi_1(x) \exp 
\Bigl(\frac{i}{2} \, \theta_{\mu\nu}\, 
\overleftarrow{\del_\mu} \, 
\overrightarrow{\del_\nu}
\Bigr) \phi_2(x)  \ .
\label{starprod_cont}
\eeq

In order to consider the lattice regularization of such theories,
we introduce a $(2m+1)$-dimensional 
toroidal lattice $\Lambda_{L,T}$ defined by
\begin{eqnarray}
\Lambda_{L,T}=\left\{ \ (x_{1},\ldots,x_{d}) 
\in \latsp \mathbb{Z}^{d} \ \Big|  \ 
-\latsp {L_\mu-1\over 2} \leq  x_\mu 
\leq \latsp {L_\mu-1\over 2} \ \right\} \ ,
\end{eqnarray}
where $\latsp$ is the lattice spacing
and $L_1 = L_2 = \cdots = L_{2m}= L$, $L_{d}= T$.
We have assumed $L,T\in \mathbb{N}$ to be odd
\cite{AMNS3}.
The dimensionful extent of the 
lattice is $\ \ell = \latsp \, L\ $ in the $2m$ spatial directions 
and $\ \tau = \latsp \, T\ $ along the Euclidean time.
The fields on the lattice are assumed to obey the periodic boundary
condition in all directions\footnote{To formulate a finite-temperature field theory,
the boundary condition in the time direction has to be taken anti-periodic for fermions.}.

In order to construct a lattice
counterpart of the star-product (\ref{starprod_cont}),
we define the Fourier transform
\beq
\tilde{\phi}(p) =
\latsp^4 \sum_{x\in \Lambda_{L,T}}
\phi(x) \,  \ee ^{- i \, p \cdot x} \ ,
\label{fourtrans}
\eeq
where the lattice momentum $p$ is discretized as
\beq
p_\mu = \frac{2 \pi n_\mu}{\latsp L_\mu}  \ ,
\quad\quad
n_\mu \in \mathbb{Z}
\ ,
\label{lattice_mom}
\eeq
and the Fourier modes $\tilde{\phi}(p)$ are periodic under
$n_\mu \mapsto n_\mu + L_\mu$.
Then the lattice star-product can be defined through its Fourier transform as
\beq
\widetilde{\phi_1 \star \phi_2}(p) 
=
\frac{1}{ \latsp ^d L^{d-1} T }
\sum_{q} \exp \left\{
- \frac{i}{2} \theta_{ij} (p-q)_i q_j \right\}
\tilde{\phi_1}(p-q) 
\tilde{\phi_2}(q)  \ ,
\label{star_lattice}
\eeq
where the noncommutativity parameter is taken to be
\beq
\theta  = 
\frac{1}{\pi} \, L \, \latsp ^2  \ .
\label{dimfultheta}
\eeq
Here and henceforth we assume that
the summation over a momentum is restricted to the Brillouin zone;
namely (\ref{lattice_mom}) 
with $-(L_\mu -1)/2 \leq  n_\mu \leq (L_\mu -1)/2 \ $.

The above lattice formulation naturally results 
from matrix model description
of noncommutative field theories, and most importantly
it preserves all the algebraic properties of the star-product.
Moreover the definition (\ref{star_lattice})
is consistent with the periodicity of the 
lattice momentum (\ref{lattice_mom}) due to (\ref{dimfultheta}).
One can also rewrite this definition (\ref{star_lattice})
in an integral form as
\beq
\phi_1(\vec{x},t) \star \phi_2(\vec{x},t)
= \frac{1}{L^{2m}} \sum _{\vec{y} }
\sum _{\vec{z} }
 \phi_1(\vec{y},t) \, \phi_2 (\vec{z},t)
\, \ee ^{-2 \, i \, (\theta ^{-1})_{ij}  \, 
( x _i - y_i ) ( x _j - z_j )} \ ,
\label{starexplicit}
\eeq
where the summation over $\vec{y}$ and $\vec{z}$ 
is taken only over the spatial lattice.
This expression is consistent with the periodicity of the
fields again due to (\ref{dimfultheta}).
As is clear from these observations,
the lattice regularization of noncommutative field 
theories inevitably requires the noncommuting directions
to be compactified in a particular way (\ref{dimfultheta}) 
consistent with the noncommutativity.
This reflects the UV/IR mixing \cite{iruv}
at a fully nonperturbative level \cite{AMNS2,AMNS3}.

Ultimately we have to take the continuum limit 
$\latsp\rightarrow 0$, and the lattice size should be sent
to infinity $L,T \rightarrow \infty$.
These two limits should be taken more carefully 
in noncommutive field theories than in commutative ones
because we have an extra scale parameter $\theta$ related
to $\latsp$ and $L$ by (\ref{dimfultheta}).
In any case we have a hierarchy of the scales
\beq
\latsp \ll \sqrt{\theta} \ll \ell
\eeq
in the regularized theory.
In order to obtain finite $\theta$,
the physical extent of the spatial direction
$\ell = \latsp L$ should inevitably go to infinity.
The extreme case $\theta \rightarrow 0$
is generally different from the commutative theory
(where $\theta = 0$ for finite $\latsp$),
as we see later in concrete examples.
The limit $T\rightarrow \infty$ in the time direction
can be taken as in commutative theories,
and one can have arbitrary $\tau$ independently of $\theta$ and
$\ell$. 

The 
U(1) gauge fields can be put on the lattice by
\beqa
U_\mu (x)  &= &{\cal P} \exp _{\star} \left( 
i g \int_x^{x +\latsp \hat{\mu}} 
{\mathcal A}_\mu (s) \, \dd s \right) \n 
&=& \sum_{n=0}^{\infty} \,  (ig)^n
\int_x^{x +\latsp \hat{\mu}}  \dd \xi_1
\int_{\xi_1}^{x +\latsp \hat{\mu}}  \dd \xi_2
\cdots 
\int_{\xi_{n-1}}^{x +\latsp \hat{\mu}}  \dd \xi_n \n 
&~& {\mathcal A}_\mu (\xi_1)  \star {\mathcal A}_\mu (\xi_2)  \star \cdots
\star {\mathcal A}_\mu (\xi_n)  \ ,
\label{U_A_rel}
\eeqa
where ${\mathcal A}_\mu(x)$ is the 
(real) gauge field in the continuum.
The path-ordering is necessary even in the U(1) case,
because of the noncommutativity arising from the star-product.
Note also that $U_\mu(x)$ is {\em not} unitary, but it is ``star-unitary'',
\beq
U_\mu(x) \star U_\mu(x)^{*} = U_\mu(x)^{*}  \star U_\mu(x) = 1 \ .
\eeq
The continuum gauge field ${\mathcal A}_\mu (x)$ transforms
under the ``star-gauge transformation'' as
\beq
{\mathcal A}_{\mu}(x)  \mapsto  g(x) 
\star {\mathcal A}_{\mu}(x) \star g(x)^{\dagger}
- \frac{i}{g} g(x) \star \frac{\partial}{\partial x_{\mu}} 
g(x)^{\dagger} \ ,
\label{star-gauge}
\eeq
where $g(x)$ is also star-unitary.
Under this transformation, the link field $U_\mu(x)$ 
defined by (\ref{U_A_rel}) transforms as
\begin{eqnarray}
U_{\mu}(x) \mapsto g(x) \star U_{\mu}(x) \star 
g(x+\latsp\hat{\mu}) ^{*} \ .
\label{trans_gauge}
\end{eqnarray}
The lattice action for the gauge field is given by
\begin{eqnarray}
S_{\rm G} = -\beta 
\sum_{x\in \Lambda_{T,N}}\sum_{\mu\neq\nu} 
U_{\mu}(x)\star U_{\nu}(x+\latsp \hat{\mu})
\star U_{\mu}(x+\latsp \hat{\nu})^{*}\star 
U_{\nu}(x)^{*} \ ,
\label{action}
\end{eqnarray}
which is invariant under star-gauge transformation (\ref{trans_gauge}).

The fermion action is defined by
\begin{eqnarray}
S_{\rm F}=
\latsp ^3 \sum_{x}\bar{\psi}(x)\star (D_{\rm w}-M) \, \psi(x) \ ,
\label{feract}
\end{eqnarray}
where $D_{\rm w}$ is the Dirac-Wilson operator
\begin{eqnarray}
D_{\rm w} = {1\over 2}\sum_{\mu=1}^{d}
\Big[\gamma_{\mu}(\nabla^{*}_{\mu}+\nabla_{\mu})
+r\latsp \nabla^{*}_{\mu}\nabla_{\mu}\Big] \ .
\label{dwop}
\end{eqnarray}
The expression of the forward and backward 
covariant derivatives depends on the transformation properties of the
fermion field. In the case where $\psi(x)$ transforms 
in the fundamental representation 
\begin{eqnarray}
\psi(s) \mapsto g(x)\star \psi(x); \hspace*{1cm} 
\bar{\psi}(x) \mapsto \bar{\psi}(x)\star g(x)^{*} \ ,
\end{eqnarray}
they are given respectively by
\begin{eqnarray}
\nabla_{\mu}\psi &=& {1\over \latsp}
\left[U_{\mu}(x)\star\psi(x+\latsp\hat{\mu})-\psi(x)\right] \nonumber \\
\nabla_{\mu}^{*}\psi &=& {1\over \latsp}\left[\psi(x)-
U_{\mu}(x-\latsp\hat{\mu})^{*}
\star \psi(x-\latsp\hat{\mu})\right] \ .
\label{covder_fund}
\end{eqnarray}
On the other hand, when fermions transform in the adjoint representation
\begin{eqnarray}
\psi(x)\mapsto g(x)\star\psi(x)\star g(x)^{*}; 
\hspace*{1cm} \bar{\psi}(x)\mapsto g(x)\star \bar{\psi}(x)\star 
g(x)^{*} \ ,
\end{eqnarray}
the forward and backward covariant derivatives are respectively defined by
\begin{eqnarray}
\nabla_{\mu}\psi &=& 
{1\over \latsp}\left[U_{\mu}(x)\star\psi(x+\latsp\hat{\mu})
\star U_{\mu}(x)^{*}
-\psi(x)\right] \nonumber \\
\nabla_{\mu}^{*}\psi &=& {1\over \latsp}
\left[\psi(x)-U_{\mu}(x-\latsp\hat{\mu})^* 
\star \psi(x-\latsp\hat{\mu})
\star U_{\mu}(x-\latsp\hat{\mu}) \right] \ .
\label{covder_adj}
\end{eqnarray}
In either case, the fermion action (\ref{feract}) is
star-gauge invariant.

The second term in the Dirac-Wilson operator (\ref{dwop}) 
is the Wilson term, which is introduced to give species doublers
a mass of order $\mathcal{O}$($1/\latsp$).
In the original proposal, the coefficient $r$ was taken to be unity,
but it can take other values, even negative ones, as far as its
magnitude is of order one.

\subsection{Feynman rules}
\label{Feynman_rules}

\begin{fmffile}{feyndiag1}

Let us proceed to formulate the perturbation theory
for the noncommutative QED on the lattice. 
As in the commutative case we start with
expanding the link variable
$U_{\mu}(x)$ in terms of the lattice gauge field $A_{\mu}(x)$ as
\begin{eqnarray}
U_\mu(x)
&=& \exp _{\star} \left\{ i\latsp gA_{\mu} (x)
 \right\} \nonumber \\
&=&
1+ig\latsp A_{\mu}(x)
-{g^2\latsp^2\over 2}
A_{\mu}(x)\star 
A_{\mu}(x) +\ldots \ .
\label{UexpandA}
\end{eqnarray}
Note that $U_\mu(x)$ is star-unitary if and only if 
$A_{\mu}(x)$ is real.
The Feynman rules are read off from the 
action (\ref{feract}) expressed in terms of the Fourier transformed
fields $\tilde{A}_{\mu}(p)$, $\tilde{\psi}(p)$ and 
$\tilde{\bar{\psi}}(p)$.
The fermion propagator is given by
\vspace*{5mm}
\begin{equation}
\parbox{20mm}{
\begin{fmfgraph*}(40,30) 
\fmfleft{i1,i2}
\fmfright{o1,o2}
\fmflabel{$p$}{v1}
\fmf{phantom}{i1,v1,i2}
\fmf{phantom}{o1,v2,o2}
\fmf{fermion,tension=0}{v1,v2}
\end{fmfgraph*}
}=
\left\{ M + \frac{1}{2} r \latsp \hat{p} ^2 
- i \gamma \cdot \tilde{p} \right\}^{-1} \equiv  Q(p)^{-1} \ ,
\label{fermprop}
\end{equation}

\vspace*{5mm}

\noindent where we have introduced the notation
\beq
\hat{p}_\mu = \frac{2}{\latsp} \sin \left(
\frac{1}{2} \latsp p_\mu \right) \ ,
\quad\quad
\tilde{p}_\mu = \frac{1}{\latsp} \sin (\latsp p_\mu) \ .
\eeq
That the fermion propagator (\ref{fermprop}) is identical to the one for the
commutative lattice QED is because the $\theta$-dependent phase arising from the star-product
(\ref{star_lattice}) disappears trivially in the quadratic term
in the action (Set $p=0$ in (\ref{star_lattice}) and consider the
anti-symmetry of $\theta_{ij}$).
The effect of the noncommutativity
will show up only in the interaction vertices in the form of
a phase depending on the momenta flowing into them.
For the one-photon vertex we find
\vspace*{1cm}
\begin{equation}
\parbox{20mm}{
\begin{fmfgraph*}(40,30) 
\fmfpen{thick}
\fmfleft{i1}
\fmfright{o1,o2}
\fmflabel{$p$}{o1}
\fmflabel{$q$}{o2}
\fmflabel{$k,\mu$}{i1}
\fmf{photon,tension=0.1}{i1,v1}
\fmf{electron,tension=0.1}{o1,v1}
\fmf{electron,tension=0.1}{v1,o2}
\end{fmfgraph*}} \hspace*{1cm}
=  \ \mathcal{W}^{(1)}(p,q) \  V_{\mu}^{(1)}(p+q) \ ,
\label{one-photon-vertex}
\end{equation}

\vspace*{1cm}

\noindent where we have defined
\beq
V_{\mu}^{(1)}(p) 
= i g \left\{
\gamma_\mu \cos \left( \frac{\latsp }{2}p_\mu\right) 
+ i r \sin \left( \frac{\latsp }{2}p_\mu\right) \right\} \ .
\label{vertexfunction1}
\eeq
The factor $\mathcal{W}^{(1)}(p,q)$ 
represents a sum of $\theta$-dependent phases,
which depends on whether the fermions couple in the fundamental or
the adjoint representation
\begin{eqnarray}
\mathcal{W}_{\rm fund}^{(1)}(p,q)&=& 
{\rm e}^{{i \over 2} {\theta }(\vec{p} \times \vec{q})} \nonumber \\
\mathcal{W}_{\rm adj}^{(1)}(p,q)&=&
{\rm e}^{{i \over 2} {\theta}(\vec{p} \times \vec{q})}
- {\rm e}^{- {i \over 2} {\theta}(\vec{p} \times \vec{q})} .
\label{W1}
\end{eqnarray}
Here we have denoted
$\vec{a}\times\vec{b}\equiv-\varepsilon_{jk}a_{j}b_{k}$, 
where $\varepsilon_{jk}$ is the matrix defined in (\ref{eps}).

For the vertex with two photons we find,

~

\begin{equation}
\parbox{20mm}{
\begin{fmfgraph*}(40,30) 
\fmfpen{thick}
\fmfleft{i1,i2}
\fmfright{o1,o2}
\fmflabel{$p$}{o1}
\fmflabel{$q$}{o2}
\fmflabel{$k,\mu$}{i1}
\fmflabel{$l,\nu$}{i2}
\fmf{photon,tension=0.1}{i1,v1}
\fmf{photon,tension=0.1}{i2,v1}
\fmf{electron,tension=0.1}{o1,v1}
\fmf{electron,tension=0.1}{v1,o2}
\end{fmfgraph*}}\hspace*{1cm}
= \ \mathcal{W}^{(2)}(p,q,k,l) \ 
V_{\mu\nu}^{(2)}(p+q)  \ ,
\end{equation}

\vspace{1cm}

\noindent where we defined
\beq
V_{\mu\nu}^{(2)}(p) 
= - \latsp g^2 \delta_{\mu\nu}
\left\{
r\cos\left({\latsp \over 2}p_\mu \right)
+ i\gamma_\mu
\sin\left(
{\latsp\over 2}p_\mu \right) \right\} \ .
\label{vertexfunction2}
\eeq
The factor $\mathcal{W}^{(2)}(p,q,k,l)$ is now given by
\begin{eqnarray}
\mathcal{W}_{\rm fund} ^{(2)}(p,q,k,l)
&=& {\rm e}^{{i \over 2} {\theta}(\vec{p}\times
\vec{q}+\vec{k}\times\vec{l})} \nonumber \\
\mathcal{W}_{\rm adj} ^{(2)}(p,q,k,l)
&=& {\rm e}^{{i \over 2}{\theta}
(\vec{p}\times\vec{q}+\vec{k}\times\vec{l})}+ 
{\rm e}^{{i \over 2} {\theta}
(-\vec{p}\times\vec{q} +\vec{k}\times\vec{l})}
- {\rm e}^{{i \over 2} {\theta}
(\vec{p}\times\vec{l}+\vec{k}\times\vec{q})}-
{\rm e}^{ {i \over 2}{\theta}
(\vec{p}\times\vec{k}+\vec{l}\times\vec{q})} \ .
\label{W2}
\end{eqnarray}
Above we assumed that the photon momenta are entering into the vertex.


In addition, for each vertex there is a Kronecker delta 
momentum conservation 
\begin{eqnarray}
\latsp ^d \, L^{d-1} \, T \, \delta_{p+k+\cdots,q} 
\end{eqnarray}
together with a summation over an internal momentum
for each loop ${1\over \latsp^d L^{d-1}T}\sum_{p}$. 
Finally, each fermion loop will carry a minus sign.
Vertices with more than two photon lines can be obtained 
in a similar way.

Together with the fermion propagator (\ref{fermprop}), the Feynman rules of noncommutative 
lattice QED also requires the photon and ghost propagators as well as the photon-ghost and photon self-interaction vertices.
In the case of the propagators, because the bilinear terms in the action are independent
of the noncommutativity parameter, they are identical to the one for ordinary QED \cite{Kawai:1980ja}.
For the interaction vertices, as it is also the case in the continuum \cite{cont}, 
they can be read off from the ones for nonabelian commutative gauge theories given
in \cite{Kawai:1980ja} 
by simply replacing the structure constants of the gauge group by the appropriate
noncommutative phases. 
In the calculation of the effective action, however, 
we will need only the fermion-photon vertices.

\subsection{Perturbative evaluation of the effective action}
\label{pert_det}

The effective action for the gauge field is defined in terms
of the fermion determinant as
\begin{eqnarray}
\Gamma[U]_{\rm eff}=-\log\left[{\det\left(D_{\rm w}-M\right)
\over \det\left(D_{\rm w,0}-M\right)}
\right] \ ,
\label{fd}
\end{eqnarray}
where $D_{\rm w,0}$ is the Dirac-Wilson operator (\ref{dwop}) 
evaluated for the trivial gauge configuration
$U_{\mu}^{(0)}(x)=\mathbf{1}$.
Expanding $U_\mu(x)$ with respect to $A_\mu (x)$ 
as in (\ref{UexpandA}),
the effective action $\Gamma[A]_{\rm eff}$ 
can be written in momentum space as
\begin{eqnarray}
\Gamma[A]_{\rm eff}
&=& {1\over 2}{1\over \latsp^d L^{d-1}T}
\sum _p
 \Pi_{\mu\nu}(p)
A_{\mu}(p)A_{\nu}(-p)
\nonumber \\
&+&  {1\over 3}{1\over (\latsp^d L^{d-1}T)^2} \sum _{p,q}
\Pi_{\mu\nu\sigma}(p,q)
A_{\mu}(p)A_{\nu}(q)A_{\sigma}(-p-q)+\ldots \ .
\label{eff_action}
\end{eqnarray}
The kernels 
$\Pi_{\mu\nu}(p)$, $\Pi_{\mu\nu\sigma}(p,q)$ 
can be computed using the diagrammatic expansion\footnote{The 
combinatorial factors in front of the diagrams take into account the 
overall factors of ${1\over 2}$ and ${1\over 3}$ 
multiplying the corresponding terms in the effective action.}.
\vspace*{5mm}
\begin{eqnarray}
\label{diag}
-\Pi_{\mu\nu}(p)&=&\hspace*{0.2cm}
\parbox{20mm}{
\begin{fmfgraph}(40,25)
\fmfkeep{2p1}
\fmfpen{thick}
\fmfleft{i}
\fmfright{o}
\fmf{photon}{i,v1}
\fmf{photon}{v2,o}
\fmf{fermion,left,tension=0.2}{v1,v2,v1}
\end{fmfgraph}
}\hspace*{-0.3cm}+
\parbox{20mm}{
\begin{fmfgraph}(40,25)
\fmfkeep{2p2}
\fmfpen{thick}
\fmfleft{i1,i2}
\fmfright{o}
\fmf{photon}{i1,v1}
\fmf{photon}{i2,v1}
\fmf{phantom}{v2,o}
\fmf{fermion,left,tension=0.2}{v1,v2,v1}
\end{fmfgraph}
}  
\label{diag_Pi2}
\\
& & \nonumber \\
-\Pi_{\mu\nu\sigma}(p,q)&=&
\parbox{20mm}{
\begin{fmfgraph}(40,25)
\fmfpen{thick}
\fmfleft{i1,i2}
\fmfright{o}
\fmf{photon}{i1,v1}
\fmf{photon}{i2,v2}
\fmf{photon}{v3,o}
\fmfcyclen{fermion,tension=0.18}{v}{3}
\end{fmfgraph}
}\hspace*{-0.3cm}
+{1\over 2}\left[
\parbox{20mm}{
\begin{fmfgraph}(40,25)
\fmfpen{thick}
\fmfleft{i1,i2}
\fmfright{o}
\fmf{photon}{i1,v1}
\fmf{photon}{i2,v1}
\fmf{photon}{v2,o}
\fmf{fermion,left,tension=0.2}{v1,v2,v1}
\end{fmfgraph}
}
\hspace*{-0.4cm}
+
\hbox{cyclic perm.}\right]
+\hspace*{0.3cm}{1\over 2}\hspace*{0.3cm}
\parbox{20mm}{
\begin{fmfgraph}(40,25)
\fmfpen{thick}
\fmfleft{i1,i2,i3}
\fmfright{o}
\fmf{photon}{i1,v1}
\fmf{photon}{i2,v1}
\fmf{photon}{i3,v1}
\fmf{phantom}{v2,o}
\fmf{fermion,left,tension=0.2}{v1,v2,v1}
\end{fmfgraph}
} 
\label{diag_Pi3}\\
~ \nonumber
\end{eqnarray}
In fact, diagrams containing vertices with three or more photons 
are irrelevant in the continuum limit since they 
are weighted with higher powers of the lattice spacing 
$\latsp$ \cite{Kawai:1980ja,MontMuns}.
Therefore we can omit the last diagram in (\ref{diag_Pi3}).
Applying the Feynman rules, we thus obtain the following expression.
\beqa
\label{Pi2_gen}
\Pi_{\mu\nu}(p)
&=&{1\over \latsp^{d}L^{d-1} T}\sum_{q}
{\rm tr\,}
\left[V_\mu^{(1)}(2q+p) \, 
Q\left(q+p\right)^{-1} \, V_\nu^{(1)}(2q+p) \, 
Q\left(q\right)^{-1} \right] \nonumber \\
&~& \, \, \times \, 
\mathcal{W}^{(1)}\left(q,q+p\right)
\, \mathcal{W}^{(1)} \left(q+{p},q\right)  
\nonumber \\
&~& + \, \, {1\over \latsp^{d}L^{d-1} T}\sum_{q}
{\rm tr\,} \left[V_{\mu\nu}^{(2)}(2q)\, Q(q)^{-1} \right] 
\mathcal{W}^{(2)}(q,q,p,-p) \ , \\
\Pi_{\mu\nu\sigma}(p_{1},p_{2})
&=& {1 \over \latsp^{d}L^{d-1} T}\sum_{q}
{\rm tr\,}
\left[V_{\mu}^{(1)}(2q+p_{1})Q\left(q+p_{1}\right)^{-1} \right. 
\nonumber \\
&~&\times \, \left. V_{\nu}^{(1)}(2q+2p_{1}+p_{2})
Q\left(q+p_{1}+p_{2}\right)^{-1}
V_{\sigma}^{(1)}(2q+p_{1}+p_{2})Q\left(q\right)^{-1}\right] 
\nonumber \\
&~& \, \, \times \, \mathcal{W}^{(1)}\left(q,q+p_{1}\right)
\mathcal{W}^{(1)}\left(q+p_{1},q+p_{1}+p_{2}\right) \nonumber \\
&~& \, \, \times \, \mathcal{W}^{(1)}\left(q+p_{1}+p_{2},q\right) \nonumber \\
&~& + \, {1 \over 2\latsp^{d}L^{d-1} T}\sum_{q} \left\{
{\rm tr\,}
\left[V_{\mu}^{(1)}(2q+p_{1})Q\left(q+p_{1}\right)^{-1}
V_{\nu\sigma}^{(2)}(2q+p_{1})Q\left(q\right)^{-1}\right] 
 \right. \nonumber \\
&~& \, \, \times \, \mathcal{W}^{(1)}\left(q,q+{p_1} \right)
\mathcal{W}^{(2)}\left(q+p_{1},q,p_{2},-p_{1}-p_{2} \right) \nonumber \\
& & \,\, +\,\, \hbox{cyclic permutations} \Big\}
  \ ,
\label{Pi3_gen}
\end{eqnarray}
where by ``cyclic permutations'' we indicate the contributions of the other two diagrams obtained from the
second one in Eq. (\ref{diag_Pi3}) by performing cyclic permutations on the labels of the external legs.
The expressions for the commutative case can be obtained
simply by omitting the factors $\mathcal{W}^{(1)}$,
$\mathcal{W}^{(2)}$ in the above equations.

\setcounter{equation}{0}
\section{The parity anomaly in 3D noncommutative QED}
\label{continuum}

From now on we will consider the three-dimensional case 
(i.e. $d\equiv 2m+1=3$) 
and study the parity anomaly in noncommutative QED on 
the lattice.
Parity anomaly has been studied intensively 
in commutative gauge theories, both 
in the continuum \cite{redlich,paranom} and on the lattice \cite{cl}. 
It has a wide application in condensed matter physics
\cite{Dunne:1998qy}
including the quantum Hall effect \cite{Khveshchenko:jh}.
We will first briefly review the known results in the commutative
case.

\subsection{A brief review of the commutative case}
\label{reviewcomm}

In three-dimensional massless QED there is a conflict between
parity symmetry and gauge invariance at the quantum level. 
As pointed out in \cite{lag-w} and elaborated in \cite{redlich},
a parity invariant regularization of the fermion determinant leads to 
non-invariance of the one-loop effective action under large gauge
transformations due to the spectral flow of the eigenvalues of the Dirac
operator, 
a phenomenon similar to the one behind Witten's global anomaly \cite{wga}.
On the other hand, a gauge invariant regularization of the theory, like
Pauli-Villars, induces a Chern-Simons action at one loop that breaks 
parity invariance, with precisely the coefficient required to compensate the 
variation of the massless fermion determinant 
under large gauge transformations.

Despite any similarities, parity anomaly in three-dimensional QED is different from ordinary 
anomalies in that the coefficient of the anomaly depends on the regularization
scheme. This peculiar aspect of parity anomaly has been
clarified by Coste and L\"uscher \cite{cl} by using
the lattice regularization, which provides the most rigid way to
calculate the anomaly while preserving gauge invariance.
Here we summarize the main results of Ref.\ \cite{cl}.
First, when $M\rightarrow 0$, one obtains in the continuum limit
\beq
\lim_{M\rightarrow 0} \Pi_{\mu\nu}(p)
={1 \over 2 \pi}
\left( n+{1\over 2}\right)
\epsilon_{\mu\nu\sigma}p_{\sigma}+
{1\over 16|p|}(p^2\delta_{\mu\nu}-p_{\mu}p_{\nu}) \ .
\label{massless}
\eeq
The term proportional to the Levi-Civita tensor is parity odd,
and hence signals the parity anomaly.
(Note that the continuum action for a massless Dirac fermion
in three dimensions is invariant under parity transformation.)
The coefficient of the parity anomaly includes 
a parameter $n$, which can take any integer value
depending on the lattice action
chosen, i.e. on the details of the ultraviolet regularization. 
The essential point, however, is that this regularization ambiguity does
not affect the existence of the parity anomaly itself, since this is 
always nonzero for any $n\in \mathbb{Z}$.
For the standard Wilson fermion, one obtains $n=0,-1$, depending on whether 
the sign of the Wilson term is positive or negative.
In this case the parity anomaly arises because the Wilson term breaks
parity on the lattice, and this breaking persists in the continuum
limit.
On the other hand, if one uses the Ginsparg-Wilson fermion, whose action is
invariant under the generalized parity transformation,
the parity anomaly (\ref{massless}) arises from the measure \cite{3dGW}, thus
realizing Fujikawa's philosophy for anomalies at a fully regularized 
level.

In the infinite mass limit, on the other hand, the result
for the vacuum polarization is given by
\beqa
\lim_{M\rightarrow + \infty} \Pi_{\mu\nu}(p)
&=&{1 \over 2 \pi} n \epsilon_{\mu\nu\sigma}p_{\sigma} 
\n
\lim_{M\rightarrow - \infty} \Pi_{\mu\nu}(p)
&=&{1 \over 2 \pi} (n+1) \epsilon_{\mu\nu\sigma}p_{\sigma} \ ,
\label{infmasslim}
\eeqa
where the integer $n$ is the same parameter as the one introduced
in (\ref{massless}) for the same lattice action.
Thus in general the fermion does not decouple
completely in the infinite mass limit but it leaves behind a certain
local term as a remnant.
In fact the freedom of the integer parameter $n$ in both
(\ref{massless}) and (\ref{infmasslim}) 
is closely related to
the fact that the remnant (\ref{infmasslim})
depends on the sign $\pm$ of the limit $M\rightarrow \pm \infty$.
By choosing different lattice action, one essentially introduces 
different numbers of heavy fermions which have masses 
of order $\mathcal{O}$($1/a$). 
The sign of the masses can be assigned as
one wishes, and this results in the arbitrariness represented by $n$.

\subsection{The noncommutative case}

Before we present our results on the noncommutative case,
let us remark on what we mean by ``parity'' when we discuss
parity anomaly in noncommutative QED.
Conventional parity refers to
a reflection in one spatial direction.
In the Euclidean formulation in three dimensions, 
one can combine the conventional
parity transformation with the 180 degrees rotation in the
remaining two directions, to arrive at the transformation
\beqa
\psi (x) &\mapsto& \psi(-x) \n
\bar{\psi} (x) &\mapsto& - \bar{\psi}(-x) \n
A_\mu (x) &\mapsto& - A_\mu(-x) \ ,
\label{mod_parity}
\eeqa
which leaves the massless Dirac action in the continuum invariant.
In the noncommutative case, the introduction of the noncommutativity
matrix $\theta_{\mu\nu}$ breaks parity in the conventional sense,
but it preserves the invariance under (\ref{mod_parity}). 
It is this invariance of the massless Dirac action
that we refer to when we say `parity anomaly' in noncommutative QED.

Our next task is to compute the effective action (\ref{fd}) 
as discussed in Section \ref{pert_det}
and to see how the results of Coste and L\"uscher \cite{cl} are modified
by noncommutative geometry.
We analyze separately the cases of fermions in the
fundamental and adjoint representation.
We will use a 
representation of the three-dimensional Dirac matrices satisfying
\begin{eqnarray}
\gamma_{\mu}\gamma_{\nu}=
\delta_{\mu\nu}+i\epsilon_{\mu\nu\sigma}\gamma_{\sigma} \ ,
\end{eqnarray}
where the matrices are taken to be Hermitian 
$\gamma_{\mu}^{\dagger}=\gamma_{\mu}$. 
Hereafter we will fix the sign of $M$ by demanding $M\geq 0$.

\subsubsection{Fundamental fermions}
\label{fund}

We begin with the coefficient
$\Pi_{\mu\nu}(p)$
of the bilinear term
in the effective action (\ref{eff_action}) for the gauge field.
For fundamental fermions the noncommutative phases 
in Eq. (\ref{Pi2_gen}) cancel out.
The resulting expression is exactly the same as in the commutative
case and 
in particular the result does not depend on how
we take the limits $L\rightarrow \infty$, $T\rightarrow \infty$ 
and $\latsp \rightarrow 0$, as far as the physical extent of the
space-time ($\ell = \latsp L$ and $\tau = \latsp T$) goes to infinity.
For instance we may take the large volume limit $L\rightarrow \infty$
and $T\rightarrow \infty$ at fixed lattice spacing $\latsp$
and then take the continuum limit $\latsp\rightarrow 0$.
Then the rest of the calculation proceeds exactly as in Ref.\ \cite{cl}.
Let us introduce the symbol
\begin{eqnarray}
T_{k}(p)f(p)=
\left.\sum_{n=0}^{k}{1\over n!}{\partial^{n}\over\partial t^{n}}f(tp)
\right|_{t=0} \ ,
\end{eqnarray} 
which represents a Taylor subtraction at zero momentum.
Thus, in the infinite volume limit, $\Pi_{\mu\nu}(p)_{\rm fund}$ 
can be rewritten as
\begin{eqnarray}
\Pi_{\mu\nu}(p)_{\rm fund}
&=& {g^2}\int_{\mathcal{B}}{d^{3}q\over (2\pi)^{3}}
\left[1-T_{0}(p)\right]{\rm tr\,}
\left[Q\left(q-{p\over 2}\right)^{-1}\partial_{\mu}Q(q)
Q\left(q+{p\over 2}\right)^{-1}\partial_{\nu}Q(q)\right]  ,
\mbox{~~~~}
\label{pi1}
\end{eqnarray}
where the large volume limit $L\rightarrow \infty$
and $T\rightarrow \infty$ has been taken and consequently
the momentum sum has been replaced by 
the integral in the Brillouin zone 
$\mathcal{B}=\{q_{\mu}\in\mathbb{R}^{3}\,|\,
-(\pi/\latsp)\leq q_{\mu}\leq (\pi/\latsp)\}$.
The subtraction of the zero external momentum contribution
comes from the tadpole diagram in the first line of Eq. (\ref{diag}).
By using the identity 
$[1-T_{0}(p)]f(p)=[1-T_{1}(p)]f(p)+p_{\mu}\partial_{\mu}f(0)$
we can write, after some algebra,
\begin{eqnarray}
\label{ff}
\Pi_{\mu\nu}(p)_{\rm fund}&=& g^2 a_{0}\epsilon_{\mu\nu\sigma}p_{\sigma}  \\
&+& {g^2}\int_{\mathcal{B}}{d^{3}q\over (2\pi)^{3}}
\left[1-T_{1}(p)\right]{\rm tr\,}
\left[Q\left(q-{p\over 2}\right)^{-1}\partial_{\mu}Q(q)\,
Q\left(q+{p\over 2}\right)^{-1}\partial_{\nu}Q(q)\right] \ , \nonumber 
\end{eqnarray}
and
\begin{eqnarray}
a_{0}={1\over 48\pi^{3}}
\int_{\mathcal{B}}d^{3}q\,\epsilon_{\mu\nu\sigma}{\rm tr\,}
\left[Q(q)^{-1}\partial_{\mu}Q(q)\,Q(q)^{-1}\partial_{\nu}Q(q)\,
Q(q)^{-1}\partial_{\sigma}Q(q)\right] \ .
\label{a0}
\end{eqnarray}
As shown in \cite{cl}, 
$a_{0}=\frac{1}{2\pi} n$ 
is a topological number, where the integer $n$ 
depends on the parameter $r$, but 
not on the lattice spacing.

Because of the subtraction at zero momentum, 
the integral in the second term on the right hand side of (\ref{ff})
has negative degree so, according to Reisz theorem \cite{reisz}, its continuum limit is given by the integral
over momentum space of the limit of the integrand when $\latsp\rightarrow 0$. On symmetry grounds, the 
two-point function in the continuum has the form
\begin{eqnarray}
\Pi_{\mu\nu}(p)=A(p)\epsilon_{\mu\nu\sigma}p_{\sigma}
+B(p)(p^2\delta_{\mu\nu}-p_{\mu}p_{\nu})
+C(p){\theta(p)_{\mu}\theta(p)_{\nu}\over 
\theta(p)^2} \ ,
\label{tpf}
\end{eqnarray}
where $\theta(p)_{\mu}\equiv \theta_{\mu\nu}p_{\nu}$. From the previous expressions, we find for the case of 
fermions in the fundamental representation
\begin{eqnarray}
A(p)_{\rm fund}&=&g^2 a_{0}+
{g^2 \over 4\pi}\int_{0}^{1}dx\left\{1-M\left[M^2+x(1-x)p^2\right]^{-{1\over 2}}\right\},
\nonumber \\
B(p)_{\rm fund} &=& {g^2 \over 2\pi}\int_{0}^{1}dx\,x(1-x)\left[M^2+x(1-x)p^2\right]^{-{1\over 2}}, \nonumber \\
C(p)_{\rm fund} &=& 0 \ .
\label{twop}
\end{eqnarray}
In the limits $M\rightarrow 0$, $M\rightarrow \infty$,
we obtain results identical to the commutative case
(\ref{massless}), (\ref{infmasslim}).

Let us proceed to 
compute the coefficient
$\Pi_{\mu\nu\sigma}(p)$
of the trilinear term
in the effective action (\ref{eff_action}) for the gauge field.
Two types of diagrams contribute to this term. In both 
cases it is easy to check that the corresponding noncommutative 
phases are identical and factor out of the
sum over the loop momentum. 
The sum multiplying the global
noncommutative phase is independent of $\theta$,
and it can be evaluated in the same limit as before.
Thus we arrive at the expression
\begin{eqnarray}
\Pi_{\mu\nu\sigma}(p_{1},p_{2})_{\rm fund} 
&=& e^{{i\over 2}\theta (\vec{p}_{1}\times\vec{p}_{2})}
\int_{\mathcal{B}}{d^{3}q\over (2\pi)^{3}}{\rm tr\,}
\left[V_{\mu}^{(1)}(2q+p_{1})Q\left(q+{p_{1}}\right)^{-1} \right.  
\label{Pi3_fund}
\\
& &\times \,\, \left. 
V_{\nu}^{(1)}(2q+2p_{1}+p_{2})Q\left(q+{p_{1}}+{p_{2}}\right)^{-1}
V_{\sigma}^{(1)}(2q+p_{1}+p_{2})Q\left(q\right)^{-1}\right] \nonumber \\
&+& {1\over 2}e^{{i\over 2}\theta(\vec{p}_{1}\times\vec{p}_{2})}\left\{
\int_{\mathcal{B}}{d^{3}q\over (2\pi)^{3}}
{\rm tr\,}\left[V_{\mu}^{(1)}(2q)Q\left(q-{p_{1}\over 2}\right)^{-1}
V_{\nu\sigma}^{(2)}(2q)Q\left(q+{p_{1}\over 2}\right)^{-1}\right]  \right.\nonumber \\
& & \,\, + \left[\mbox{cyclic permutations:}\,\, (\mu,p_{1}) \rightarrow (\nu,p_{2}) \rightarrow (\sigma,-p_{1}-p_{2})
\right]\Big\} \ ,
\nonumber
\end{eqnarray}
where we have taken the large volume limit $L\rightarrow \infty$
and $T\rightarrow \infty$, but the continuum limit is yet to be taken.
In general, the three-point function can be written as
\begin{eqnarray}
\Pi_{\mu\nu\sigma}(p_{1},p_{2})_{\rm fund}
={\rm e}^{{i\over 2}\theta(\vec{p}_{1}\times\vec{p}_{2})}\mathcal{A}(p_{1},p_{2})_{\rm fund}\epsilon_{\mu\nu\sigma}
+\ldots \ ,
\label{fg2p}
\end{eqnarray}
where ``$\ldots$'' stands for terms proportional to rank-three tensors constructed in terms of the incoming momenta.
By looking at the low momentum expansion we find that
\begin{eqnarray}
\mathcal{A}(p_{1},p_{2})_{\rm fund} = -g^{3} a_{0}+\mathcal{O}(p^2),
\end{eqnarray}
where $a_{0}$ is given again by Eq. (\ref{a0}). The remaining momentum-dependent contribution can be obtained by 
evaluating 
\begin{eqnarray}
& & \int_{\mathcal{B}}{d^{3}q\over (2\pi)^{3}}
\left[1-T_{0}(p_{1},p_{2})\right]{\rm tr\,}
\left[V_{\mu}^{(1)}(2q+p_{1})Q\left(q+{p_{1}}\right)^{-1}
V_{\nu}^{(1)}(2q+2p_{1}+p_{2})\right. \nonumber \\
& & \hspace*{1cm} \times \,\,\left. Q\left(q+{p_{1}}+{p_{2}}\right)^{-1}
V_{\sigma}^{(1)}(2q+p_{1}+p_{2})Q\left(q\right)^{-1}\right] 
\label{ff3}
\end{eqnarray}
and keeping the terms proportional to $\epsilon_{\mu\nu\sigma}$.
Here we have denoted $T_{0}(p,q)f(p,q)=f(0,0)$. 
Note also that the second group of terms in (\ref{Pi3_fund}) does not 
contribute to $\mathcal{A}(p_{1},p_{2})_{\rm fund}$, since they are symmetric in two of the indices.

In order to retrieve the continuum limit of (\ref{ff3}) we need to check that the relevant terms of the 
integral on the right-hand side converges
to the continuum Feynman integral. By writing the integrand as $V(q,p_{1},p_{2},M,a)/C(q,p_{1},p_{2},M,a)$ we find 
that ${\rm deg\,}V\leq 8$, whereas ${\rm deg\,}C=12$, so the degree of divergence of the integrand is 
$3+{\rm deg\,}V-{\rm deg\,}C\leq -1$, and the contribution of the integral in the continuum limit is given by
the integral over the loop momentum of the $\latsp\rightarrow 0$ limit of the integrand. A careful evaluation of the
resulting integral using Feynman parameters shows that (see Appendix \ref{A2} for the details)
\begin{eqnarray}
\lim_{a\rightarrow 0}\mathcal{A}(p_{1},p_{2})_{\rm fund} 
&=&-g^{3}a_{0}-{g^{3}\over 2\pi}\int_{0}^{1}dx_{1}\int_{0}^{1-x_{1}}dx_{2}
\left[1-M(M^2+\Delta)^{-{1\over 2}}\right] \nonumber \\
&-&{g^{3}\over 4\pi}\,M\,p_{1}^{2}\int_{0}^{1}dx_{1}
\int_{0}^{1-x_{1}}dx_{2}{(x_{1}+x_{2})(1-x_{1}-x_{2})\over
(M^2+\Delta)^{3\over 2}} \nonumber \\
&-&{g^{3}\over 4\pi}\,M \,p_{2}^2\,
\int_{0}^{1}dx_{1}\int_{0}^{1-x_{1}}dx_{2}\,{x_{2}(1-x_{2}) \over 
(M^2+\Delta)^{3\over 2} } 
\label{threep}\\
&-&{g^3\over 8\pi}\,M\,p_{1}\cdot p_{2}
\int_{0}^{1}dx_{1}\int_{0}^{1-x_{1}}dx_{2}\,{x_{1}+2x_{2}(1-x_{1}-x_{2})
\over (M^2+\Delta)^{3\over 2}} \ , \nonumber 
\nonumber 
\end{eqnarray}
where for shorthand we have written
\begin{eqnarray}
\Delta\equiv (x_{1}+x_{2})(1-x_{1}-x_{2})p_{1}^2
+x_{2}(1-x_{2})p_{2}^{2}+2x_{2}(1-x_{1}-x_{2})p_{1}\cdot p_{2} \ .
\label{delta}
\end{eqnarray}
From these results we see that
the cubic term in the noncommutative Chern-Simons action emerges 
in the limits $M\rightarrow 0$
and $M\rightarrow  \infty$ with coefficients consistent with
the quadratic term calculated above. 
This is expected from the star-gauge invariance,
which is manifestly preserved
in the lattice regularization.

It is straightforward to calculate now the effective action $\Gamma[A]_{\rm eff}^{\rm (fund)}$ in position space for 
massless fermions. By using Eqs. (\ref{twop}) and (\ref{threep}) in the limit $M\rightarrow 0$
we find for the parity violating part
\begin{eqnarray}
\Gamma[A]_{\rm eff}^{\rm (fund)}
&=& {g^2\over 4\pi}\left(n+{1\over 2}\right)\int {d^{3}p\over (2\pi)^3}\epsilon_{\mu\nu\sigma}
\tilde{A}_{\mu}(p)\tilde{A}_{\nu}(p)p_{\sigma} \\
&-& {g^3\over 6\pi}\left(n+{1\over 2}\right)\int{d^3 p_{1}\over (2\pi)^3}\int{d^3 p_{1}\over (2\pi)^3}
\epsilon_{\mu\nu\sigma}\tilde{A}_{\mu}(p_{1})\tilde{A}_{\nu}(p_{2})\tilde{A}_{\sigma}(-p_{1}-p_{2})
{\rm e}^{{i\over 2}\theta(\vec{p}_{1}\times \vec{p}_{2})}. \nonumber 
\end{eqnarray}
Performing the inverse Fourier transform on $\tilde{A}_{\mu}(p)$ and defining $\mathcal{A}_{\mu}(x)=ig A_{\mu}(x)$ we 
finally arrive at
\begin{eqnarray}
\Gamma[\mathcal{A}]_{\rm eff}^{\rm (fund)}&=& {i\over 4\pi}\left(n+{1\over 2}\right)\int d^{3}x\,\epsilon_{\mu\nu\sigma}\,
\left(\mathcal{A}_{\mu}\partial_{\nu}\mathcal{A}_{\sigma}+
{2\over 3}\mathcal{A}_{\mu}\star \mathcal{A}_{\nu}\star \mathcal{A}_{\sigma}\right)
\label{nccs}
\end{eqnarray}
In particular, since
the NCCS term is not invariant under the parity transformation (\ref{mod_parity}),
we have a parity anomaly
as in the commutative case. 

It is important to notice that in the $\theta \rightarrow 0 $ limit we retrieve 
the results obtained in Ref. \cite{cl} for ordinary (commutative) QED.
The fact that the commutative limit
turned out to be smooth in the present case
is due to the cancellation of the noncommutative phases involving
loop momenta, which would otherwise cause the UV/IR mixing.
Such a cancellation can be understood in a transparent way \cite{iruv}
if one uses the so-called double-line notation known from large-$N$ gauge
theory. Feynman rules should be re-derived accordingly and 
in particular each of the interaction vertices will have a {\em single}
noncommutative phase factor instead of a sum of phases.
Usefulness of the double-line notation
in noncommutative field theories can be understood
if one recalls that
the algebraic property of the star-product is the same as that
of matrix product.
In the double-line notation, ``planar diagrams'' can be defined
as diagrams which can be drawn on a plane without any crossings
of lines.
In fact for any planar diagrams
the noncommutative phase factors out of the momentum integration,
leaving a global phase depending only on external momenta 
\cite{Gonzalez-Arroyo:1983ac,Filk:dm,Ishibashi:1999hs}.
For fermions in the fundamental representation,
the interaction with the gauge field occurs only on one side
of the fermion propagator (represented as a double-line), 
and therefore all the diagrams that appear in the calculation of 
the effective action are actually planar.


\subsubsection{Adjoint fermions}
\label{adj}

Contrary to the case studied above, 
the effective action for noncommutative gauge theories 
with fermions in the adjoint representation does not have 
a smooth commutative limit.
Note that the original theory in the $\theta_{\mu\nu}\rightarrow 0$
limit is just a free massless fermion and a free photon, 
which is parity invariant. 
However, if we take the continuum limit
for finite $\theta$ we will see that the adjoint fermions induce
the NCCS action, which reduces to an ordinary (commutative) 
U(1) Chern-Simons term in the $\theta_{\mu\nu}\rightarrow 0$ limit.
Using a Pauli-Villars fermion of mass $M_{\rm reg}$ as a regulator in the continuum \cite{Chu} 
this is the result of the fact that the two limits
$\theta_{\mu\nu}\rightarrow 0$ and 
$M_{\rm reg}\rightarrow  \infty$ 
do not commute with each other, a phenomenon characteristic of 
noncommutative quantum field theories both at zero \cite{iruv} and finite
temperature \cite{avm}.
This comes about because of the existence of nonplanar diagrams in which
the UV divergences are regulated at the scale\footnote{The role of 
nonplanar amplitudes in the calculation of the chiral anomaly has attracted some
attention recently (see, for example, \cite{nonplanar,alt}).} $\theta|\vec{p}|$.

Plugging in the factors $\mathcal{W}^{(1)}$,
$\mathcal{W}^{(2)}$ for the adjoint fermions (\ref{W1}), (\ref{W2}),
the expression (\ref{Pi2_gen}) 
for the two-point function becomes
\begin{eqnarray}
\Pi_{\mu\nu}(p)_{\rm adj}&=&2\Pi_{\mu\nu}(p)_{\rm fund} \nonumber \\
& &\hspace*{-2cm} -{2\over \latsp^{3} L^2 T}\sum_{q}{\rm tr\,}\left[
V_{\mu}^{(1)}(2q+p)Q\left(q+{p}\right)^{-1}V_{\nu}^{(1)}(2q+p)\,
Q\left(q\right)^{-1}\right]
\cos\left[\theta(\vec{p}\times\vec{q})\right] \nonumber \\
& &\hspace*{-2cm}-{2\over \latsp^3 L^2 T}\sum_{q}{\rm tr\,}
\left[V_{\mu\nu}^{(2)}(2q)
Q(q)^{-1}\right]\cos\left[\theta(\vec{p}\times \vec{q})\right] \ ,
\label{Pi2_adj}
\end{eqnarray}
where $\Pi_{\mu\nu}(p)_{\rm fund}$ is 
the two-point function for fermions in the
fundamental representation calculated above. 
In the language of the double-line notation 
(See the end of Sec.\ \ref{fund}), 
the first term represents
the contribution from the planar diagram and the other two terms correspond
to the nonplanar contributions.
Interestingly the third term, which is the non-planar contribution 
from the tadpole diagram,
exactly vanishes in the continuum limit.
Therefore the effect of the tadpole diagram 
is just to subtract the zero-momentum contribution 
from the planar terms,
which has the effect of making the amplitude 
finite in the continuum limit, as seen above.
Note that the second term of (\ref{Pi2_adj})
is finite by itself,
since the noncommutative phase introduces an effective
cutoff to the loop momentum at the scale $\theta |\vec{p}|$.

To obtain the continuum limit of the second term
we have to take into account that
this limit has to be taken at the same time 
with the large volume limit in 
such a way that $\theta \propto \latsp^2 L$ is kept fixed. 
A long but straightforward calculation shows that the resulting 
integrals can be written in terms of
modified Bessel functions of the second kind. 
The resulting two-point function 
has the form given in Eq. (\ref{tpf}) now with (see Appendix \ref{A2} for details)
\begin{eqnarray}
A(p)_{\rm adj}&=&
2A(p)_{\rm fund}+{g^2\over 2\pi}\sqrt{2\over \pi}|\theta(\vec{p})|^{1\over 2}\int_{0}^{1}dx\,\Big\{
M[M^2+x(1-x)p^2]^{-{1\over 4} } \nonumber \\
& & \times\,\, K_{1\over 2}\left(\theta|\vec{p}|[M^2+x(1-x)p^2]^{1\over
2}\right)\Big\} ,
\nonumber \\
B(p)_{\rm adj} &=& 
2B(p)_{\rm fund}-{g^2\over \pi}\sqrt{2\over \pi}|\theta(\vec{p})|^{1\over 2}\int_{0}^{1}dx\,\Big\{
\,x(1-x)[M^2+x(1-x)p^2]^{-{1\over 4}} \nonumber \\
& & \times \,\, K_{1\over 2}\left(\theta|\vec{p}|[M^2+x(1-x)p^2]^{1\over
2}\right)\Big\} ,
\nonumber \\
C(p)_{\rm adj}&=& 
{g^2\over \pi}\sqrt{2\over \pi}|\theta(p)|^{1\over 2}\int_{0}^{1}dx \Big\{ \left[M^2+x(1-x)p^2\right]^{3\over 4}
\nonumber \\ 
& & \times \,\,
K_{3\over 2}\left(\theta|\vec{p}|[M^2+x(1-x)p^2]^{1\over
2}\right)\Big\} \ .
\label{soladj2p}
\end{eqnarray}
Using the asymptotic expansion 
for large arguments of the Bessel functions, 
$K_{\nu}(z)\sim e^{z}/\sqrt{2\pi z}$ 
it is easy to see that the nonplanar terms
in the amplitude are exponentially suppressed in the limit of large
$M$ (cf. \cite{Chu}). 
In the $M\rightarrow 0$ limit, on the other hand, one finds that
the nonplanar contributions to the parity violating term vanishes.
Thus in both limits
we find that the parity violating term in the effective action 
comes solely from the planar part, and its magnitude is 
twice the one for fundamental fermions. Later we will offer a physical
explanation of this phenomenon.

For finite values of the noncommutative parameter $\theta$ the three-dimensional
Euclidean group SO(3) is broken down to SO(2), acting as rotations on the noncommutative
plane.
An important consequence of the smooth commutative limit ($\theta\rightarrow 0$)
of noncommutative QED with fundamental fermions is that the full Euclidean group 
SO(3) is restored in that limit. This means that the ``generalized parity'' anomaly
studied here is mapped to the usual parity anomaly of commutative QED.
On the other hand, for adjoint fermions the situation
is radically different, since the Euclid-breaking term in the effective action, third 
equation in  
(\ref{soladj2p}),
does not disappear in the commutative limit and actually induces divergences when 
$\theta\rightarrow 0$, while keeping  the SO(2) symmetry unbroken (see the discussion at
the end of this subsection). The reason is that in this case, because of UV/IR mixing, the
noncommutative theory is not a smooth deformation of (free) commutative QED with
neutral fermions.

Let us now turn to the evaluation of the three-point function. 
After a short manipulation of the 
noncommutative phases, 
the three-point function can be written as
\begin{eqnarray}
\Pi_{\mu\nu\sigma}(p_{1},p_{2})_{\rm adj}
&=&\left[\left({\rm e}^{{i\over 2}\theta(\vec{p}_{1}\times\vec{p}_{2})}
-{\rm e}^{-{i\over 2}\theta(\vec{p}_{1}\times\vec{p}_{2})}\right)
\mathcal{A}(p_{1},p_{2})_{\rm fund}+
\mathcal{A}(p_{1},p_{2},\theta)_{\rm NP}\right]\epsilon_{\mu\nu\sigma}
\nonumber \\
&+&\ldots \ ,
\label{ppt}
\end{eqnarray}
where $\mathcal{A}(p_{1},p_{2})_{\rm fund}$ is the same function appearing in Eq.  (\ref{fg2p}) and 
$\mathcal{A}(p_{1},p_{2},\theta)_{\rm NP}$ is
the nonplanar contribution. As above the ellipsis denotes further terms whose tensor structure depends on 
the external momenta. The nonplanar function $\mathcal{A}(p_{1},p_{2})_{\rm NP}$ is calculated by 
evaluating the sum
\begin{eqnarray}
& & {1\over \latsp^{3}L^2T}\sum_{q}
{\rm tr\,}
\left[V_{\mu}^{(1)}(2q+p_{1})\,Q\left(q+{p_{1}}\right)^{-1}V_{\nu}^{(1)}(2q+2p_{1}+p_{2}) \right. \\
& & \left. \hspace*{1cm}\times\,\, 
Q\left(q+{p_{1}}+{p_{2}}\right)^{-1} 
V_{\sigma}^{(1)}(2q+p_{1}+p_{2})\,Q\left(q\right)^{-1} \right]
\mathcal{W}(p_{1},p_{2},q)_{\rm NP} \nonumber 
\end{eqnarray}
and keeping the terms proportional to $\epsilon_{\mu\nu\sigma}$. Here the 
``nonplanar" part of the noncommutative phase $\mathcal{W}(p_{1},p_{2},q)_{\rm NP}$ is
given by
\begin{eqnarray}
\mathcal{W}(p_{1},p_{2},q)_{\rm NP} &=& 
{\rm e}^{{i\over 2}\theta(\vec{p}_{1}\times\vec{p}_{2})}\left[{\rm e}^{i\theta(\vec{q}\times\vec{p}_{2})}
-{\rm e}^{-i\theta(\vec{q}\times\vec{p}_{1})}
-{\rm e}^{i\theta\vec{q}\times(\vec{p}_{1}+\vec{p}_{2})}\right] \nonumber \\
&-& 
{\rm e}^{-{i\over 2}\theta(\vec{p}_{1}\times\vec{p}_{2})}\left[{\rm e}^{-i\theta(\vec{q}\times\vec{p}_{2})}
-{\rm e}^{i\theta(\vec{q}\times\vec{p}_{1})}
-{\rm e}^{-i\theta\vec{q}\times(\vec{p}_{1}+\vec{p}_{2})}\right] 
\ .
\label{npph}
\end{eqnarray}
$\mathcal{A}(p_{1},p_{2})_{\rm NP}$ can be 
computed in the continuum limit $\latsp\rightarrow 0$
at infinite volume with
$a^2 L$ fixed and expressed in terms of the modified Bessel functions of the second
kind. In this limit, $\mathcal{A}(p_{1},p_{2},\theta)_{\rm NP}$ is expressed in terms of sum of integrals of 
the form
\begin{eqnarray}
g^{3}M\int{d^{3}q\over (2\pi)^{3}}{M^2+q^2+(p_{1}+p_{2})\cdot q\over
[M^2+(q+p_{1})^2][M^2+(q+p_{1}+p_{2})^2](M^2+q^2)} {\rm e}^{i\theta(\vec{w}\times \vec{q})},
\label{archetype}
\end{eqnarray}
where $\vec{w}=\alpha \,\vec{p}_{1}+\beta \, \vec{p}_{2}$ ($\alpha,\beta=0,\pm 1$) 
is a linear combination of the incoming momenta, that can be read off from Eq. (\ref{npph}).
These integrals can be computed again in terms of modified Bessel functions of the second kind. 
As it was the case also for the two-point functions, all the terms contributing to 
$\mathcal{A}(p_{1},p_{2},\theta)_{\rm NP}$ 
vanish both in the limit $M\rightarrow 0$, due to the global factor of $M$ in front of the integral, and
as $M\rightarrow \infty$, this time due to the exponential damping of the modified Bessel function for large values of
the argument. The final conclusion is that in the continuum limit
\begin{eqnarray}
\lim_{M\rightarrow 0} \mathcal{A}(p_{1},p_{2},\theta)_{\rm NP} = \lim_{M\rightarrow \infty} 
\mathcal{A}(p_{1},p_{2},\theta)_{\rm NP} =0.
\end{eqnarray}
As a consequence, in these two limits, the coefficient of the Chern-Simons action 
is only determined by the first term on the right-hand side of (\ref{ppt}) 
and its value is consistent with the one calculated above
from the two-point function. This, again, follows from the fact that the lattice regularization
preserves star-gauge invariance.

We can now study the parity-odd part of the
induced effective action in the limit $M\rightarrow 0$. From the two- and three-point amplitudes calculated above we find
that $\Gamma[\mathcal{A}]_{\rm eff}^{\rm (adj)}=2\Gamma[\mathcal{A}]_{\rm eff}^{\rm (fund)}$ 
where $\Gamma[\mathcal{A}]_{\rm eff}^{\rm (fund)}$
is given in Eq. (\ref{nccs}). Note that we have considered that the adjoint fermions are of Dirac type. The minimal form of
the anomaly, however, is obtained by imposing the Majorana condition. In this case there is an extra 
factor of ${1\over 2}$ in front of the fermionic determinant and the effective action for Majorana fermions in
the adjoint representation agrees with Eq. (\ref{nccs}) \cite{Chu}.

To conclude, let us try to understand in physical terms the vanishing of the nonplanar contribution to the parity-violating
part of the effective action in the limit of zero fermion mass. As mentioned in Section \ref{reviewcomm} the parity anomaly
in commutative gauge theories results from the impossibility of finding a parity-invariant UV cutoff which 
at the same time preserves Lorentz and gauge symmetries. This is clear from the analysis of \cite{redlich}
where the introduction of a parity-invariant UV
cutoff $\Lambda$ in the integrals results in the presence of a term $g^2 \Lambda \eta_{\mu\nu}$ in the two-point function
which breaks the Ward identity. 

Because of the presence of the noncommutative phases depending on the loop momentum, we can see the nonplanar contribution 
to the two-point function of {\it massless} noncommutative QED as a regularization of the corresponding amplitude in ordinary 
(commutative) massless QED. In this case the noncommutative momentum $\theta|\vec{p}|\equiv \Lambda^{-1}$ plays the role 
of an UV cutoff. Moreover, this cutoff preserve the parity invariance of the theory. Therefore one expects that
the parity-breaking terms in the amplitudes will vanish. Indeed, using the nonplanar part in (\ref{soladj2p}) we find
the two-point function of massless QED in this regularization to be
\begin{eqnarray}
\Pi_{\mu\nu}(p)^{\theta}_{{\rm QED}}= {g^2\over 2\pi|p|^{1\over 2}}\sqrt{2\over\pi}\Lambda^{-{1\over 2}}
(p^2\delta_{\mu\nu}-p_{\mu}p_{\mu})
\int_{0}^{1}dx
[x(1-x)]^{{3\over 4}}K_{1\over 2}\left(\Lambda^{-1}[x(1-x)p^2]^{1\over 2}\right) \nonumber \\
-\,\,{g^2\over 2\pi}\sqrt{2\over \pi}|p|^{3\over 2}\Lambda^{-{1\over 2}}{\theta(p)_{\mu}\theta(p)_{\nu}\over
\theta(p)^2}\int_{0}^{1} dx\,[x(1-x)]^{3\over 4}K_{3\over 2}\left(\Lambda^{-1}[x(1-x)p^2]^{1\over 2}\right),\hspace*{2cm}
\label{regQED}
\end{eqnarray}
i.e. the two-point function does not contain parity-breaking terms and satisfies the Ward identity.
However, because of the presence of the last term, it breaks Euclidean symmetry. Thus the ``$\theta$-regularization'' 
of massless QED
provides a regularization scheme in which both parity and gauge symmetries are maintained at the cost of breaking
Euclidean (or Lorentzian) invariance. In the limit in which the 
cutoff is sent to infinity, $\Lambda\rightarrow
\infty$, the coefficient of the first term in the right-hand side of (\ref{regQED}) tends to $g^2/(16|p|)$ whereas the 
second Euclid-breaking term diverges linearly with $\Lambda$.
Euclidean invariance can be restored by introducing a Pauli-Villars fermion
with mass $M_{\rm reg}$ which subtracts the divergent part of the two-point function (\ref{regQED}). However, 
this procedure results in the breaking of parity invariance and as a consequence a Chern-Simons term is again
induced in the limit $M_{\rm reg}\rightarrow\infty$. This is somewhat reminiscent 
to the analysis of nonplanar anomalies in noncommutative gauge theories presented in 
\cite{alt}.

\setcounter{equation}{0}
\section{Noncommutative Chern-Simons theory on the lattice}
\label{CSterm}

In this section our results in the previous sections are
used to define a lattice-regularized
noncommutative Chern-Simons theory
following the proposal made in Ref.\ \cite{3dGW} in the 
commutative case.
The basic idea is to use the parity breaking part of the
effective action induced by the Ginsparg-Wilson fermion.
Since masslessness of the Ginsparg-Wilson fermion is guaranteed
on the lattice, one obtains the correct
noncommutative Chern-Simons action in the continuum limit
without fine-tuning.




The Dirac operator $D$ for the Ginsparg-Wilson fermion
is characterized by
the Ginsparg-Wilson relation \cite{GW}
which, in the form applicable to both even and odd dimensions,
is given by \cite{3dGW}
\beq
D+D^{\dagger}=\latsp D^{\dagger}D \ .
\label{ggr}
\eeq
The general solution to Eq.\ (\ref{ggr}) can be written as
\beq
D = \frac{1}{\latsp} (1 - V) \ ,
\label{overlapDirac}
\eeq
where $V$ is a unitary operator, which should turn into 
the identity operator in the naive continuum limit.
In even dimensions, assuming further the 
``$\gamma_{5}$-Hermiticity''
\beq  \label{gamma5H}
D^{\dag} = \gamma_{5} D \gamma_{5} \ ,
\eeq
one arrives at the original Ginsparg-Wilson relation \cite{GW}
\beq \label{GWR5}
D \gamma_{5} + \gamma_{5} D = a \, D \gamma_{5} D \ .
\eeq
This relation guarantees that the fermion action
including the operator $D$ is invariant
under a generalized chiral symmetry \cite{ML},
which reduces to the ordinary chiral symmetry
in the continuum limit.
The role of the ``$\gamma_{5}$-Hermiticity''
is played in odd dimensions by the property
\beq
D (U) ^{\dag}
=   {\cal R } D (U^P) {\cal R }  \ ,
\label{reflect}
\eeq
where $U^{P}$ is the parity transformed gauge configuration,
\beq
U_{\mu}^{P}(x) = U_{\mu}(-x)^{\dag} \ ,
\eeq
and ${\cal R}$
is the space-time reflection operator,
${\cal R} : x \mapsto - x $.
Combining (\ref{ggr}) and (\ref{reflect}) one can
show the invariance of the corresponding fermion action
under a generalized parity transformation \cite{3dGW}.
The measure, however, is not invariant under the same transformation.
As a consequence, the fermion determinant is not invariant 
but transforms as
\beq
\det D(U^P) = (\det V)^* \det D(U) \ .
\label{detD_UP}
\eeq

So far we have discussed general properties of the 
Ginsparg-Wilson operator, which satisfies (\ref{ggr}).
In fact the unitary operator $V$ has to be chosen appropriately 
in order to guarantee that the operator $D$ has 
sensible properties as a Dirac operator
such as locality (with exponentially decaying tails)
and the absence of species doublers.
Such an operator has been derived from the overlap formalism
\cite{NN}, and it is given explicitly by \cite{overlapDirac}
\beqa
\label{defV}
V &=& A_{\rm w} / \sqrt{ A_{\rm w} ^ \dag A_{\rm w} }  \\
A_{\rm w} &=&  1 - \latsp D_{\rm w}(r=-1)  \ ,
\label{defAw}
\eeqa
where $D_{\rm w}(r=-1)$ is the Dirac-Wilson operator,
which has the form (\ref{dwop}) with ordinary covariant derivatives.
The noncommutative version of the Ginsparg-Wilson fermion
can be obtained by simply using the covariant derivatives
(\ref{covder_fund}) or (\ref{covder_adj}) 
depending on the representation,
instead of the usual ones without star-products.
In even dimensions Ginsparg-Wilson fermions 
played a crucial role in introducing chirality 
on a discretized noncommutative torus \cite{Nishimura:2001dq}.
Recently an analogous construction has been worked out
on a fuzzy sphere \cite{Aoki:2002fq}.

For the choice (\ref{defV}), $\det V$ in (\ref{detD_UP})
is nothing but the phase of $\det A_{\rm w}$,
which is essentially the fermion determinant of the
Wilson-Dirac operator with $r=-1$ and $M=1/\latsp$.
Thus one can translate the result obtained for 
the Wilson fermion in the infinite mass limit
into the parity anomaly for the Ginsparg-Wilson fermion.
In the commutative case this is how the correct parity anomaly
has been reproduced by \cite{Kikukawa:1998qh}
in the overlap formalism\footnote{See also \cite{Narayanan:1997by} for an
earlier work on the overlap formalism in odd dimensions, where a parity invariant
phase choice has been made.}.
In the noncommutative case, on the other hand,
our results in the previous section with $r=-1$ 
in the limit $M\rightarrow \infty$
implies that the parity anomaly
obtained for Ginsparg-Wilson fermions coincides with the 
result for Wilson fermions with $r=-1$ 
in the massless limit $M\rightarrow 0$.



As in the commutative case \cite{3dGW},
the parity anomaly for the Ginsparg-Wilson fermion 
suggests a natural definition of the noncommutative
Chern-Simons term on the lattice.
Namely we define it as $S_{\rm CS}$ in
\beq
\ee ^{i S_{\rm CS}} \defeq 
\frac{\det A_{\rm w}}{|\det A_{\rm w}|}  \ ,
\label{latticeCS}
\eeq
where $A_{\rm w}$ is defined by (\ref{defAw}) with
the covariant derivative (\ref{covder_fund})
for the fundamental representation.
Here we remind the reader that $A_{\rm W}$ is related to 
the {\em infinite-mass} Dirac determinant as we discussed 
in the previous paragraph.
According to our calculations,
the quantity $S_{\rm CS}$
indeed becomes the noncommutative Chern-Simons action
in the continuum limit.
In the continuum, on the other hand, 
noncommutative Chern-Simons term is known to transform as 
\cite{lqnccs}\footnote{This gauge violation was also concluded in Ref.\ \cite{sjcs}, 
although it was overlooked in its first version.}
\beq
S_{\rm CS} \mapsto S_{\rm CS} + 2 \pi \nu \ ,
\eeq
under a gauge transformation, where $\nu$ is the winding number
characterizing this gauge transformation.
The gauge invariance requires
the coefficient of the noncommutative Chern-Simons action
to be quantized.
That Eq.\ (\ref{latticeCS}) defines $S_{\rm CS}$ only up to
modulo $2 \pi$ is therefore not a problem for most practical purposes. 
Note in this regard that the right-hand side of eq.\
(\ref{latticeCS}) is indeed manifestly gauge invariant.

\setcounter{equation}{0}
\section{Concluding remarks}
\label{summary}

In the present paper we have studied the emergence of parity anomaly 
on the lattice for
three-dimensional noncommutative QED, both with fermions in the
fundamental and the adjoint 
representation. Induced Chern-Simons actions in noncommutative gauge
theories have been studied in the continuum in \cite{Chu,MS}
using the Pauli-Villars regularization (see also \cite{NCCSref} for 
an incomplete list of references). 
However, the main advantage of the lattice analysis
presented here lies in making explicit the dependence 
of the coefficient of the induced Chern-Simons term 
on the regularization scheme used. 
Thus, the results obtained in Ref.\ \cite{Chu} 
corresponds to the cases $n=0,-1$ in our analysis 
using the lattice regularization. Notice that
the quantization of the scheme-dependent term 
in the effective action is consistent with the
star-gauge invariance of the fermion determinant 
under ``large'' transformations \cite{lqnccs}, 
as required by the fact that the lattice regularization 
respects star-gauge invariance.

We have also proposed a lattice-regularized 
Chern-Simons action on a noncommutative torus
using Ginsparg-Wilson fermions.
As the lattice formulation of noncommutative field theories
has been useful to extract their interesting nonperturbative
dynamics \cite{2dNCYM,Bietenholz:2002vj,Ambjorn:2002nj},
we hope that the lattice formulation
of noncommutative Chern-Simons action is useful
to deepen our understanding of quantum Hall systems.

Finally we would like to emphasize that the lattice 
noncommutative field theories studied in the present paper
can be mapped on to a finite $N$ matrix model.
The anomaly calculation in matrix models has recently 
attracted attention
in the context of large-$N$ gauge theory 
\cite{Kiskis:2002gr,Kikukawa:2002ms}
and noncommutative geometry \cite{Aoki:2002mn,Aoki:2002fq}.
We expect that the calculation method developed
in this paper is useful to study various anomalies in 
noncommutative geometry.
In particular we would like to revisit the gauge anomaly cancellation
in chiral gauge theories on a noncommutative torus
\cite{Nishimura:2001dq}.
We hope that such developments will eventually lead 
us to a deeper understanding
of the stringy nature of the space-time structure.

\end{fmffile}

\bigskip

\acknowledgments

We would like to thank L. Alvarez-Gaum\'e, A. Armoni, J.L.F.~Barb\'on, 
M.~Garc\'{\i}a-P\'erez, S.~Iso, H.~Kawai,
Y.~Kikukawa, K.E.~Kunze and D.~Miller for helpful discussions. 
The work of J.N.\ is supported in part by Grant-in-Aid for 
Scientific Research (No.\ 14740163) from 
the Ministry of Education, Culture, Sports, Science and Technology.
M.A.V.-M. acknowledges partial support from Spanish
Science Ministry Grants AEN99-0315 and FPA2002-02037.

\bigskip

\appendix
\section{Functional determinants}
\label{a1}

In the following we will provide an alternative calculation of the effective action $\Gamma[A]_{\rm eff}$ on the lattice
by direct evaluation of the fermionic determinant (\ref{fd}). From the definition of the Dirac-Wilson operator 
(\ref{dwop}) and the expansion of the link field $U_{\mu}(x)$ in terms of the lattice gauge field $A_{\mu}(x)$ 
(\ref{UexpandA}), one can write
\begin{eqnarray}
D_{\rm w}=\sum_{k=0}^{\infty} g^{k}D_{{\rm w},k}.
\label{expansionD}
\end{eqnarray}
Using this expansion, the effective action can be expressed as
\begin{eqnarray}
\Gamma[A]_{\rm eff}&=&-\log\left[{\det{(D_{\rm w}-M)}\over \det{(D_{{\rm w},0}-M)}}\right] = -\log\det{\left[ 
1+(D_{{\rm w},0}-M)^{-1}\sum_{k=1}^{\infty}g^{k}D_{{\rm w},k}\right]} \nonumber \\
&=& -{\rm Tr\,}\log\left[1+\sum_{k=1}^{\infty}g^{k}(D_{{\rm w},0}-M)^{-1}D_{{\rm w},k}\right],
\end{eqnarray}
which leads to the following series for $\Gamma[A]_{\rm eff}$:
\begin{eqnarray}
\Gamma[A]_{\rm eff} = g^2\Gamma_{2}[A]+g^{3}\Gamma_{3}[A]+\ldots
\end{eqnarray}
with
\begin{eqnarray}
\Gamma_{2}[A]&=& {\rm Tr\,}\left\{ {1\over 2}\left[(D_{{\rm w},0}-M)^{-1}D_{{\rm w},1}\right]^{2}-(D_{{\rm w},0}-M)^{-1}D_{\rm w,2}
\right\} 
\label{g2} \\
\Gamma_{3}[A] &=& {\rm Tr\,}\Big\{-{1\over 3}\left[(D_{{\rm w},0}-M)^{-1}D_{{\rm w},1}\right]^{3}+
\left[(D_{{\rm w},0}-M)^{-1}D_{{\rm w},1}(D_{{\rm w},0}-M)^{-1}D_{{\rm w},2}\right] \nonumber \\
& & \,\,-\,\,(D_{{\rm w},0}-M)^{-1}D_{{\rm w},3}\Big\}.
\label{g3}
\end{eqnarray}
By comparing with Eqs. (\ref{diag_Pi2}) and (\ref{diag_Pi3}) we can identify each term in the previous equations
with the contribution of a particular Feynman diagram. 

A quick computation shows that for fermions in the fundamental representation the operator $D_{{\rm w},k}$ 
appearing in Eq. (\ref{expansionD}) is given by
\begin{eqnarray}
D_{{\rm w},k}^{({\rm fund})}\psi(x)&=&{(i\latsp)^{k}\over 2\latsp k!}\sum_{\mu=1}^{d}\left[
(r+\gamma_{\mu})A_{\mu}(x)^{\,\star k}\star \psi(x+\latsp\hat{\mu}) \right.\nonumber \\
& & \,\,+\,\,\left. (-1)^{k}(r-\gamma_{\mu})
A_{\mu}(x-\latsp\hat{\mu})^{\,\star k}\star\psi(x-\latsp\hat{\mu})\right],
\end{eqnarray}
whereas when the fermions are in the adjoint representation the result is
\begin{eqnarray}
D_{{\rm w},k}^{({\rm adj})} \psi(x)
&=& {(i\latsp)^{k}\over 2\latsp k!}\sum_{\mu=1}^{d}\sum_{m=0}^{k}(-1)^{m}
\left(
\begin{array}{c}
k \\
m
\end{array}
\right)
\left[(r+\gamma_{\mu})A_{\mu}(x)^{\,\star(k-m)}\star\psi(x+\latsp\hat{\mu})\star A_{\mu}(x)^{\,\star m} \right. \nonumber \\
& & +\,\,\left. (r-\gamma_{\mu})A_{\mu}(x-\latsp\hat{\mu})^{\,\star m}\star\psi(x-\latsp\hat{\mu})\star 
A_{\mu}(x-\latsp\hat{\mu})^{\,\star (k-m)}\right].
\end{eqnarray}
In both expressions we have used the notation $\phi(x)^{\,\star n}\equiv \overbrace{\phi(x)\star\ldots\star\phi(x)}^{n}$.

In order to evaluate each term in Eq. (\ref{g2})-(\ref{g3}) it is convenient to 
work in momentum space. In the following we will detail the calculation for fundamental fermions, 
leaving the adjoint case for the reader. Using 
Eq. (\ref{fourtrans}) together with
\begin{eqnarray}
\tilde{A}_{\mu}(p) = \latsp^{3} \sum_{x\in\Lambda_{L,T}}A_{\mu}(x){\rm e}^{-ip\cdot\left(x+{1\over 2}\latsp\hat{\mu}\right)}
\end{eqnarray}
one can easily find the action of the operators in Eqs. (\ref{g2})-(\ref{g3}) 
on $\tilde{\psi}(p)$. For the free propagator we have
\begin{eqnarray}
{(D_{{\rm w},0}-M)^{-1}\tilde{\psi}}(p)&=& -Q(p)^{-1}\tilde{\psi}(p),
\end{eqnarray}
whereas the result for $D_{{\rm w},1}$ can be written as
\begin{eqnarray}
g{D_{1}\tilde{\psi}}(p) &=& {ig\over \latsp^{3} L^2 T}\sum_{q\in\mathcal{B}}\sum_{\mu=1}^{3}\left\{
\gamma_{\mu}\cos\left[{\latsp\over 2}(p+q)_{\mu}\right]+ir\sin\left[{\latsp\over 2}(p+q)_{\mu}\right]\right\} 
{\rm e}^{{i\over 2}\theta(\vec{p}\times \vec{q})}
\nonumber \\
& & \times \,\,\tilde{A}_{\mu}(p-q)\tilde{\psi}(q)
\nonumber \\
 &= & {1\over  \latsp^{3} L^2 T}\sum_{q\in\mathcal{B}}\sum_{\mu=1}^{3}V_{\mu}^{(1)}(p+q)
\mathcal{W}_{\rm fund}^{(1)}(p,q)
\tilde{A}_{\mu}(p-q)\tilde{\psi}(q).
\end{eqnarray}
Here $V_{\mu}^{(1)}(p)$ and $\mathcal{W}_{\rm fund}^{(1)}(p,q)$ are defined in Eqs. (\ref{vertexfunction1}) and (\ref{W1})
respectively.
For $D_{{\rm w},2}$ we arrive at
\begin{eqnarray}
g^2{D_{{\rm w},2}\tilde{\psi}}(p) &=& -{\latsp g^2\over 2(\latsp^{3} L^2 T)^2}\sum_{q,q'\in \mathcal{B}}
\sum_{\mu=1}^{3}\left\{
r\cos\left[{\latsp\over 2}(p+q)_{\mu}\right]+i\gamma_{\mu}\sin\left[{\latsp\over 2}(p+q)_{\mu}\right]\right\}
 \nonumber \\
& &\times \,\, {\rm e}^{{i\over 2}\theta\left[\vec{p}\times\vec{q}+\vec{q}\,'\times (\vec{p}-\vec{q}-\vec{q}\,')\right]}
\tilde{A}_{\mu}(q')\tilde{A}_{\mu}(p-q-q')\tilde{\psi}(q)
\nonumber \\
 &=& {1\over  2(\latsp^{3} L^2 T)^2}\sum_{q,q'\in\mathcal{B}}\sum_{\mu,\nu=1}^{3}
V_{\mu\nu}^{(2)}(p+q)\mathcal{W}^{(2)}_{\rm fund}(p,q,q',p-q-q') \nonumber \\
& & \times \,\,\tilde{A}_{\mu}(q')\tilde{A}_{\nu}(p-q-q')\tilde{\psi}(q).
\end{eqnarray}
As in the previous case we have introduced the vertex function and the noncommutative phase defined in Eqs. (\ref{vertexfunction2})  
and (\ref{W2}) respectively.
Finally, for $D_{{\rm w},3}$ the result is
\begin{eqnarray}
g^3{D_{{\rm w},3}\tilde{\psi}}(p) &=& {\latsp g^3\over 3!(\latsp^{3} L^2 T)^3}\sum_{q,q',q''\in \mathcal{B}}
\sum_{\mu=1}^{3}\left\{
r\sin\left[{\latsp\over 2}(p+q)_{\mu}\right]-i\gamma_{\mu}\cos\left[{\latsp\over 2}(p+q)_{\mu}\right]\right\}
 \nonumber \\
& & \times \,\, {\rm e}^{{i\over 2}\theta\left[\vec{q}\times\vec{q}\,'+\vec{q}\times\vec{q}\,''+\vec{q}\,'\times\vec{q}\,''+
\vec{p}\times(\vec{q}-\vec{q}\,'-\vec{q}\,'')\right]} \nonumber \\
& &\times\,\,\tilde{A}_{\mu}(q')\tilde{A}_{\mu}(q'')\tilde{A}_{\mu}(p-q-q'-q'')\tilde{\psi}(q) .
\end{eqnarray}
This term is associated with the three-photon vertex in the diagramatic expansion that, as argued in Section \ref{Feynman_rules},
is irrelevant in the continuum limit.

We have seen that, when written in momentum space, all operators appearing in Eqs. (\ref{g2})-(\ref{g3}) are expressed as finite matrices, 
$\mathcal{O}\tilde{\psi}(p)=\sum_{q\in\mathcal{B}}\mathcal{O}(p,q)\tilde{\psi}(q)$, whose
traces can be easily calculated. Let us begin with $\Gamma_{2}[A]$. The first trace to be computed is
\begin{eqnarray}
g^2{\rm Tr\,}\left\{ {1\over 2}\left[(D_{{\rm w},0}-M)^{-1}D_{{\rm w},1}\right]^{2}\right\} &=& 
{1\over 2(\latsp^{3} L^2 T)^2}\sum_{\mu,\nu=1}^{3}\sum_{p,q\in\mathcal{B}}\tilde{A}_{\mu}(p)\tilde{A}_{\nu}(-p) \nonumber \\
\hspace*{-2cm}&\hspace*{-2cm} &\hspace*{-2cm}
\times \,
{\rm tr\,}\left[V_{\mu}^{(1)}(2q+p)Q\left(q+p\right)^{-1}V_{\nu}^{(1)}(2q+p)Q\left(q\right)^{-1}\right],
\label{g11}
\end{eqnarray}
where ``tr'' indicates the trace over Dirac indices. In the same way, for the second 
trace in (\ref{g2}) the result is
\begin{eqnarray}
-g^2{\rm Tr\,}\left[(D_{{\rm w},0}-M)^{-1}D_{\rm w,2}\right] &=& 
{1\over 2(\latsp^{3} L^2 T)^2}\sum_{\mu,\nu=1}^{3}\sum_{p,q\in\mathcal{B}}\tilde{A}_{\mu}(p)\tilde{A}_{\nu}(-p) 
\nonumber \\
& & \times\,{\rm tr\,}\left[V_{\mu\nu}^{(2)}(2q)\,Q(q)^{-1}\right].
\label{g12}
\end{eqnarray}
Adding these two terms to get $\Gamma_{2}[A]$ and extracting the kernel $\Pi_{\mu\nu}(p)$ defined in (\ref{eff_action}), 
we recover Eq. (\ref{Pi2_gen}). 

The cubic term $\Gamma_{3}[A]$ in the effective action can be computed along similar lines. The first term in (\ref{g3})
gives
\begin{eqnarray}
-{1\over 3}{\rm Tr\,}& &\!\!\!\left\{\left[(D_{{\rm w},0}-M)^{-1}D_{{\rm w},1}\right]^{3}\right\} ={1\over 3(\latsp^{3}L^2 T)^3}
\sum_{\mu,\nu,\sigma=1}^{3}\,\,\sum_{p,q,q'\in\mathcal{B}} \tilde{A}_{\mu}(q)\tilde{A}_{\nu}(q')\tilde{A}_{\sigma}(-q-q')
 \nonumber \\
& & \times \,\,{\rm e}^{{i\over 2}\theta(\vec{q}\times\vec{q}\,')} \,{\rm tr\,}
\left[V_{\mu}^{(1)}(2p+q)Q\left(p+{q}\right)^{-1} 
V_{\nu}^{(1)}(2p+2q+q')
Q\left(p+{q}+{q'}\right)^{-1} \right. \nonumber \\
& & \left. \times \,\,V_{\sigma}^{(1)}(2p+q+q')Q\left(p\right)^{-1}\right] ,
\label{uno}
\end{eqnarray}
whereas the second trace renders
\begin{eqnarray}
{\rm Tr}& &\!\!\!\!\!\!\! \left[(D_{{\rm w},0}-M)^{-1}D_{{\rm w},1}(D_{{\rm w},0}-M)^{-1}D_{{\rm w},2}\right] \nonumber \\
&=& 
{1\over 2(\latsp^{3}L^2 T)^{3} } 
\sum_{\mu,\nu,\sigma=1}^{3}\,\,\sum_{p,q,q'\in\mathcal{B}} \tilde{A}_{\mu}(q)\tilde{A}_{\nu}(q')\tilde{A}_{\sigma}(-q-q') \nonumber \\
&\times & {\rm e}^{{i\over 2}\theta(\vec{q}\times\vec{q}\,')} \,{\rm tr\,}
\left[V_{\mu}^{(1)}(2p+q)Q\left(p+q\right)^{-1}V_{\nu\sigma}^{(2)}(2p+q)\,Q\left(p+q\right)^{-1}\right].
\label{dos}
\end{eqnarray}
The third term ${\rm Tr\,}\left\{[D_{{\rm w},0}-M]^{-1}D_{{\rm w},3}\right\}$ corresponds to the contribution of the
tadpole diagram which is irrelevant in the continuum limit. Adding together (\ref{uno}) and (\ref{dos}), and identifying
the kernel $\Pi_{\mu\nu\sigma}$, we recover the result of Eq. (\ref{Pi3_gen}).

In the case of adjoint fermions, the calculation is analogous to the one describe above, the main difference being the 
noncommutative phases. Again the results of Section \ref{pert_det} are recovered.

\section{Evaluation of the Feynman integrals}
\label{A2}

In this Appendix we will provide the reader with details of the calculation of some of the 
Feynman integrals in Sections \ref{fund} and \ref{adj}.
As explained above, we consider a hermitian representation of the $2\times 2$ gamma matrices 
satisfying $\gamma_{\mu}\gamma_{\nu}=\delta_{\mu\nu}+i\epsilon_{\mu\nu\sigma}\gamma_{\sigma}$. This implies the
following trace identities:
\begin{eqnarray}
{\rm tr\,}(\gamma_{\mu}\gamma_{\nu}\gamma_{\sigma})&=& 2i\epsilon_{\mu\nu\sigma}\,, \nonumber \\
{\rm tr\,}(\gamma_{\mu}\gamma_{\nu}\gamma_{\sigma}\gamma_{\lambda}) &=& 2\left(\delta_{\mu\nu}\delta_{\sigma\lambda}
+\delta_{\mu\lambda}\delta_{\nu\sigma}-\delta_{\mu\sigma}\delta_{\nu\lambda}\right), \nonumber \\
{\rm tr\,}(\gamma_{\mu}\gamma_{\nu}\gamma_{\sigma}\gamma_{\lambda}\gamma_{\alpha}) &=& 
2i\left(\delta_{\mu\nu}\epsilon_{\sigma\lambda\alpha}+\delta_{\lambda\alpha}\epsilon_{\mu\nu\sigma}+
\delta_{\sigma\lambda}\epsilon_{\mu\nu\alpha}-\delta_{\sigma\alpha}\epsilon_{\mu\nu\lambda}\right), \nonumber \\
{\rm tr\,}(\gamma_{\mu}\gamma_{\nu}\gamma_{\sigma}\gamma_{\lambda}\gamma_{\alpha}\gamma_{\beta}) &=& 
2\left(\delta_{\mu\nu}\delta_{\sigma\lambda}\delta_{\alpha\beta}
+\delta_{\mu\nu}\delta_{\sigma\beta}\delta_{\lambda\alpha}
-\delta_{\mu\nu}\delta_{\sigma\alpha}\delta_{\lambda\beta}
-\delta_{\alpha\beta}\delta_{\mu\sigma}\delta_{\nu\lambda} 
\right. \nonumber \\
&+& \left. \delta_{\alpha\beta}\delta_{\mu\lambda}\delta_{\nu\sigma} 
-\delta_{\lambda\alpha}\delta_{\mu\sigma}\delta_{\nu\beta}
+ \delta_{\lambda\alpha}\delta_{\mu\beta}\delta_{\nu\sigma}
+\delta_{\lambda\beta}\delta_{\mu\sigma}\delta_{\nu\alpha}  
\right. \nonumber \\
&-& \left. \delta_{\lambda\beta}\delta_{\mu\alpha}\delta_{\nu\sigma}-\epsilon_{\mu\nu\sigma}\epsilon_{\lambda\alpha\beta}\right).
\label{trid}
\end{eqnarray}

\subsection{Fundamental fermions}

As shown in \cite{cl}, the integrand of the second term in Eq. (\ref{ff}) has negative degree so the continuum limit
exists and gives rise to the integral
\begin{eqnarray}
I_{\mu\nu}=-g^2 \int {d^{3}q\over (2\pi)^{3}}\left[1-T_{1}(p)\right]{{\rm tr\,}\left\{\gamma_{\mu}\left[M+i(\fsl{q}+
\fsl{p})\right]\gamma_{\nu}\left[M+i\fsl{q}\right]\right\}\over \left[M^2+(q+p)^2\right]\left(M^2+q^2\right)}.
\end{eqnarray}
By using the trace identities (\ref{trid}) and writing the denominator as an integral over a Feynman parameter
\begin{eqnarray}
{1\over \left[M^2+(q+p)^2\right]\left(M^2+q^2\right)}=\int_{0}^{1}dx\left[(q+xp)^2+M^2+x(1-x)p^2\right]^{-2},
\end{eqnarray}
one arrives at
\begin{eqnarray}
I_{\mu\nu}&=&-2g^2\int_{0}^{1}dx\int{d^{3}q\over (2\pi)^{3}}\left[1-T_{1}(p)\right]\Big\{M\epsilon_{\mu\nu\alpha}p_{\alpha}
-2q_{\mu\nu}+2x(1-x)p_{\mu}p_{\nu}
\nonumber \\
& & \,+\,\, \left[M^2+q^2-x(1-x)p^2\right]\delta_{\mu\nu}\Big\}\left[q^2+M^2+x(1-x)p^2\right]^{-2}.
\label{finalint}
\end{eqnarray}
The final result (\ref{twop}) is readily obtained by computing the momentum integral. Note that, because of the zero-momentum
subtraction, the integral in (\ref{finalint}) is free of divergences.

We now evaluate the function $\mathcal{A}(p_{1},p_{2})_{\rm fund}$ in (\ref{fg2p}). The relevant integral to calculate is the 
continuum limit of Eq. (\ref{ff3}) which can be cast into
\begin{eqnarray}
I_{\mu\nu\sigma}&=&-ig^{3}\int{d^{3}q\over (2\pi)^{3}}\left[1-T_{0}(p_{1},p_{2})\right] \nonumber \\
& &\times\,\,{{\rm tr}\Big\{\gamma_{\mu}\left[M+i(\fsl{q}+\fsl{p}_{1})\right]\gamma_{\nu}
\left[M+i(\fsl{q}+\fsl{p}_{1}+\fsl{p}_{2})\right]\gamma_{\sigma}(M+i\fsl{q})\Big\}\over
\left[M^2+(q+p_{1})^{2}\right]\left[M^2+(q+p_{1}+p_{2})^{2}\right](M^2+q^2)}.
\label{a2}
\end{eqnarray}
In order to compute $\mathcal{A}(p_{1},p_{2})_{\rm fund}$ we need to retain only those terms proportional to the 
Levi-Civita tensor $\epsilon_{\mu\nu\sigma}$. By expanding the trace in the numerator and using Eqs. (\ref{trid})
it is straightforward to check that only two terms, proportional to $M^{3}$ and  $M$, 
contribute to $\mathcal{A}(p_{1},p_{2})_{\rm fund}$, namely 
\begin{eqnarray}
\mathcal{A}(p_{1},p_{2})_{\rm fund}&=& 4Mg^{3}\int{d^{3}q\over (2\pi)^{3}}\int_{0}^{1}dx_{1}\int_{0}^{1-x_{2}}dx_{2}
\left[1-T_{0}(p_{1},p_{2})\right] \nonumber \\
& &\times\,\, {M^2+q^2+(p_{1}+p_{2})\cdot q\over \Big\{ \left[q+(x_{1}+x_{2})p_{1}+x_{2}p_{2}\right]^2+\Delta\Big\}^{3}},
\label{i3planar}
\end{eqnarray}
where we have reduced the denominator of (\ref{a2}) by introducing Feynman parameters and $\Delta$ is defined in 
Eq. (\ref{delta}). The integral can be easily evaluated using
standard techniques to find the result given in Eq. (\ref{threep}).

\subsection{Adjoint fermions}

In the case when the fermions are in the adjoint representation, the evaluation of the Feynman integrals is more
involved due to the presence of noncommutative phases dependent on the loop momentum. Let us focus first on the 
two-point function. As explained above, contrary to the planar part where
the tadpole diagram introduces the zero momentum subtraction that makes the whole amplitude finite, in the 
nonplanar sector the tadpole diagram cancels exactly. This can be seen by noticing that, 
up to a total derivative term that cancels in the continuum limit, the nonplanar contribution of the tadpole diagram
in the continuum limit is given by
\begin{eqnarray}
\Pi_{\mu\nu}(p,\theta)_{\rm adj}\Big|_{\rm tadpole} &=& -4g^2\int {d^{3}q\over (2\pi)^{3}}{(M^2+q^2)\delta_{\mu\nu}
-2q_{\mu}q_{\nu} \over (q^2+M^2)^2}\cos\left[\theta(\vec{p}\times\vec{q})\right] \nonumber \\
&+& 4g^2\theta(p)_{\nu}\int{d^{3}q\over (2\pi)^{3}}{q_{\mu}\over q^2+M^2}\sin\left[\theta(\vec{p}\times\vec{q})\right].
\label{tadpole}
\end{eqnarray}
The relevant integrals can be easily solved in terms of modified Bessel functions of the second kind ($\omega\in\mathbb{R}$):
\begin{eqnarray}
\int {d^{3}q\over (2\pi)^{3}} {2q_{\mu}q_{\nu}-q^2\delta_{\mu\nu} \over
(q^2+\omega^2)^2} {\rm e}^{i\theta(\vec{p}\times\vec{q})} &=& {1\over 8\pi}\sqrt{2\over \pi}
|\omega|^{3\over 2}(\theta |\vec{p}|)^{1\over 2}\Big\{
K_{1\over 2}\left(\theta|\vec{p}|\,|\omega|\right)
\delta_{\mu\nu} \nonumber \\
& & -\,\,2K_{3\over 2}\left(\theta|\vec{p}|\,|\omega|\right){\theta(p)_{\mu}\theta(p)_{\nu}
\over \theta(p)^2} \Big\} ,
\label{i1} \\
\int{d^{3}q\over (2\pi)^{3}}{{\rm e}^{i\theta(\vec{p}\times\vec{q})}\over (q^2+\omega^2)^2} &=& {1\over 8\pi}\sqrt{2\over \pi}
|\omega|^{-{1\over 2}}(\theta|\vec{p}|)^{1\over 2}K_{1\over 2}\left(\theta|\vec{p}|\,|\omega|\right) ,
\label{i2} \\
\int {d^{3}q\over (2\pi)^{3}}{q_{\mu}\over q^2+\omega^2}{\rm e}^{i\theta(\vec{p}\times\vec{q})}&=& 
{i\over 4\pi}{\theta(p)_{\mu}\over \theta(p)^2}\sqrt{2\over\pi}(\theta|\vec{p}|)^{1\over 2}|\omega|^{{3\over 2}}
K_{3\over 2}\left(\theta|\vec{p}|\,|\omega|\right).
\label{i3}
\end{eqnarray}
Substituting these expressions into (\ref{tadpole}) one finds a cancellation between the different terms.

As for the nonplanar part of the two-point function coming from the first diagram in Eq. (\ref{diag}), the relevant integral
to evaluate is
\begin{eqnarray}
I_{\mu\nu}^{\theta}=-g^2 \int {d^{3}q\over (2\pi)^{3}}{{\rm tr\,}\left\{\gamma_{\mu}\left[M+i(\fsl{q}+
\fsl{p})\right]\gamma_{\nu}\left[M+i\fsl{q}\right]\right\}\over \left[M^2+(q+p)^2\right]\left(M^2+q^2\right)}
\,{\rm e}^{i\theta(\vec{p}\times\vec{q})}.
\end{eqnarray}
Here one can follow the same steps as in the case of fundamental fermions, leading to
\begin{eqnarray}
I_{\mu\nu}^{\theta}&=&-2g^2\int_{0}^{1}dx\int{d^{3}q\over (2\pi)^{3}}  \\
& & \times \,\, {M\epsilon_{\mu\nu\alpha}p_{\alpha}-2q_{\mu\nu}+2x(1-x)p_{\mu}p_{\nu} 
+ \left[M^2+q^2-x(1-x)p^2\right]\delta_{\mu\nu}  \over
\left[q^2+M^2+x(1-x)p^2\right]^{2}}\,{\rm e}^{i\theta(\vec{p}\times\vec{q})}.  
\nonumber 
\end{eqnarray}
Once more, by using (\ref{i1})-(\ref{i3}) one readily finds the expressions (\ref{soladj2p}).

To conclude, we outline the calculation of the function
$\mathcal{A}(p_{1},p_{2},\theta)_{\rm NP}$ in Eq. (\ref{ppt}).
As in the case of the two point function, the only difference with respect to the planar part analyzed in the previous 
subsection [cf. Eq. (\ref{i3planar})] is the presence of the noncommutative phase. This yields
\begin{eqnarray}
\mathcal{A}(p_{1},p_{2},\theta)_{\rm NP}&=& 4Mg^{3}\int{d^{3}q\over (2\pi)^{3}}\int_{0}^{1}dx_{1}\int_{0}^{1-x_{2}}dx_{2} \nonumber \\
& & \times \,\,
{M^2+q^2+(p_{1}+p_{2})\cdot q\over \Big\{ \left[q+(x_{1}+x_{2})p_{1}+x_{2}p_{2}\right]^2+\Delta\Big\}^{3}}
\mathcal{W}(p_{1},p_{2},q)_{\rm NP},
\label{i3nonplanar}
\end{eqnarray}
where the phases are given in Eq. (\ref{npph}). Because of the structure of the noncommutative 
phases, $\mathcal{A}(p_{1},p_{2},\theta)_{\rm NP}$ is indeed
a sum of terms of the form (\ref{archetype}). After shifting the 
loop momentum, the integral can be evaluated with the help of Eq. (\ref{i2}) together with
\begin{eqnarray}
\int{d^{3}q\over (2\pi)^{3}}{q_{\mu}{\rm e}^{i\theta(\vec{p}\times\vec{q})}\over (q^2+\omega^2)^{3}} &=& 
{i\over 32\pi}{\theta(p)_{\mu}\over \theta(p)^2}\sqrt{2\over \pi}|\omega|^{-{1\over 2}}(\theta|\vec{p}|)^{5\over 2}
K_{1\over 2}\left(\theta|\vec{p}|\,|\omega|\right), \nonumber \\
\int{d^{3}q\over (2\pi)^{3}}{{\rm e}^{i\theta(\vec{p}\times\vec{q})}\over (q^2+\omega^2)^{3}} &=&
{1\over 32\pi}\sqrt{2\over \pi}(\theta|\vec{p}|)^{3\over 2}|\omega|^{-{3\over 2}}
K_{3\over 2}\left(\theta|\vec{p}|\,|\omega|\right).
\end{eqnarray}
In our case the constant $\omega$ is replaced by $\Delta$. Thus, for large values of the fermion mass
and at fixed incoming
momenta the argument of the Bessel functions is very large and the corresponding integrals vanish exponentially. 
In the same way, if $M\rightarrow 0$ at finite momenta $\Delta$ is nonzero and the corresponding integrals remain
finite. Thus, because of the presence of an overall power of $M$ in front of (\ref{i3nonplanar}), 
all the integrals contributing to the function
$\mathcal{A}(p_{1},p_{2},\theta)_{\rm NP}$ will vanish in that limit.

\end{document}